\documentclass[english]{IEEEtran}
\usepackage[T1]{fontenc}
\usepackage[latin9]{inputenc}
\usepackage{color}
\usepackage{babel}
\usepackage{verbatim}
\usepackage{float}
\usepackage{textcomp}
\usepackage{amsmath}
\usepackage{amsthm}
\usepackage{amssymb}
\usepackage{graphicx}
\usepackage[unicode=true]
 {hyperref}

\makeatletter

\floatstyle{ruled}
\newfloat{algorithm}{tbp}{loa}
\providecommand{\algorithmname}{Algorithm}
\floatname{algorithm}{\protect\algorithmname}
\usepackage{tikz}
\usetikzlibrary{calc}

\theoremstyle{plain}
\newtheorem{thm}{\protect\theoremname}
\theoremstyle{definition}
\newtheorem{defn}[thm]{\protect\definitionname}
\theoremstyle{plain}
\newtheorem{lem}[thm]{\protect\lemmaname}
\theoremstyle{plain}
\newtheorem{cor}[thm]{\protect\corollaryname}
\theoremstyle{plain}
\newtheorem{prop}[thm]{\protect\propositionname}

\usepackage{cite}
\usepackage{tikz,pgf}
\usetikzlibrary{calc,arrows}
\usepackage{algorithm,algpseudocode}

\makeatother

\providecommand{\corollaryname}{Corollary}
\providecommand{\definitionname}{Definition}
\providecommand{\lemmaname}{Lemma}
\providecommand{\propositionname}{Proposition}
\providecommand{\theoremname}{Theorem}

\begin{document}
\title{Improved Rate-Energy Trade-off For SWIPT Using Chordal Distance Decomposition
In Interference Alignment Networks}
\author{Navneet Garg\href{https://orcid.org/0000-0001-8535-7663}{\includegraphics[width=0.3cm]{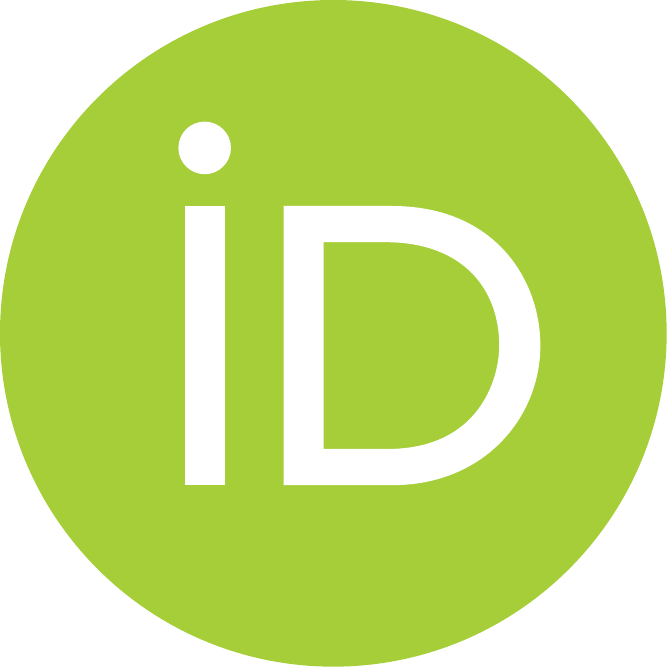}},
Avinash Rudraksh, Govind Sharma, Tharmalingam Ratnarajah\thanks{N. Garg and T. Ratnarajah are with The University of Edinburgh, UK.
A. Rudraksha and G. Sharma are with Indian Institute of Technology
Kanpur, India. E-mails: \{ngarg@ed.ac.uk, avinash.rudraksha@gmail.com,
govind@iitk.ac.in, t.ratnarajah@ed.ac.uk\}. This work was supported
by the UK Engineering and Physical Sciences Research Council (EPSRC)
under grant number EP/P009549/1.} }

\maketitle

\begin{abstract}
This paper investigates the simultaneous wireless information and
power transfer (SWIPT) precoding scheme for K-user multiple-input-multiple-output
(MIMO) interference channels (IC), for which interference alignment
(IA) schemes provide optimal precoders to achieve full degrees-of-freedom
(DoF) gain. However, harvesting RF energy simultaneously reduces the
achievable DoFs. To study a trade-off between harvested energy and
sum rate, the transceiver design problem is suboptimally formulated
in literature via convex relaxations, which is still computationally
intensive, especially for battery limited nodes running on harvested
energy. In this paper, we propose a systematic method using chordal
distance (CD) decomposition to obtain the balanced precoding, which
improves the trade-off. Analysis shows that given the nonnegative
value of CD, the achieved harvested energy for the proposed precoder
is higher than that for perfect IA precoder. Moreover, energy constraints
can be achieved, while maintaining a constant rate loss without losing
DoFs via tuning the CD value and splitting factor. Simulation results
verify the analysis and add that the IA schemes based on max-SINR
or mean-squared error are better suited for SWIPT maximization than
subspace or leakage minimization methods. 
\end{abstract}

\begin{IEEEkeywords}
Chordal distance; interference alignment; power splitting; rate-energy
trade-off; Simultaneous wireless information and power transfer (SWIPT).
\end{IEEEkeywords}

\section{\label{sec:Introduction}Introduction}

In wireless networks, energy consumption is one of the major issues
due to increasing number of devices and the need for environment protection
\cite{7120020}. Green communications have attracted much interest
from academia and industry. In the past few years, wireless energy
harvesting (EH) has emerged as an important method to achieve green
wireless communications \cite{Wan2011362}. In EH, the energy collected
from the ambient environment can be utilized as a power supply for
self-sustained wireless nodes \cite{Prasad2014195,Valenta2014108}.
Since radio frequency (RF) signals carry energy, these signals can
act as a new source for EH. Wireless power transfer (WPT) is becoming
an important segment in future wireless communications. The experiment
results in \cite{Vyas20132491} demonstrate that a few microwatts
of RF power is harvested from the broadcast signals of TV stations,
which are located at several kms away. Therefore, wireless EH system
can be used for energy-constrained devices, smart wearables, and implantable
sensors \cite{Krikidis2014104}. On the other hand, since RF signals
also communicate information in wireless systems, simultaneous information
and power transfer (SWIPT) technology have attracted great research
interests \cite{Khandaker2014277,Shi20143269,Shi20146194,Xu2014,Zong2016430,8254370}.
Some pioneering works on SWIPT have been done in \cite{Varshney2008,Grover2010},
where the rate-energy region has been characterized for single antenna
point-to-point system. In \cite{Zhang2013}, for multiple-input-multiple-output
(MIMO) broadcast network, dedicated EH and information decoding (ID)
receivers are used. The time switching duration (in time-division
access) or power splitting (PS) ratio (in frequency-division access)
is computed using iterative convexified algorithm in \cite{Liu20133990}.
\cite{Chen2014407} proposes the quantized CSI feedback to improve
EH, and derives the trade-off between EH duration and ID.

Further, for interference channels (ICs), interference is one of the
most fundamental and challenging aspects. Regarding interference cancellation,
from the last decade, interference alignment (IA) has emerged as a
promising solution for MIMO wireless networks. Under certain conditions,
IA has been shown to be degree of freedom (DoF) optimal for ICs \cite{Cadambe20083425}.
In IA, the precoders at the sources and the decoders at the destinations
are employed to align and cancel the interfering signal from other
users \cite{Cadambe20083425}. To design these IA precoders and decoders,
several iterative algorithms have been investigated in the literature,
such as signal-to-interference-plus-noise-ratio (SINR) maximization,
leakage minimization \cite{Garg2016finite,Xu2013b}, mean squared
error (MSE) minimization \cite{Razavi2015,Garg2016d2d}, alternating
minimization \cite{Peters20092445}, etc. These IA algorithms assume
channel state information (CSI) available at the sources to compute
the IA precoders. In frequency division duplexed (FDD) systems, this
information is obtained at the sources using CSI feedback either in
quantized or in analog form \cite{Anand2013,Ravindran2008,Jindal2006,Krishnamachari2013,8379356}.
In quantized CSI feedback, the linear rate scaling is maintained at
a given signal-to-noise-ratio (SNR), only if the number of bits are
scaled proportional to SNR \cite{Krishnamachari2013,garg2015precoder,Garg2016quant,Chen2014}.
For analog feedback, a constant rate-loss has been observed at medium-to-high
SNR regime, i.e., without any loss of DoFs. 

Next, for IA networks, the interference component is canceled at each
receiver to separate out the desired signal. However, before interference
nulling, the received signal can be split, and the interference power
can be utilized for harvesting energy. A review of SWIPT schemes is
given as follows. In \cite{Park2013591} with $2$-users IC, different
possible transmission strategies are defined for time-switching (TS)
receivers. Authors in \cite{Lee0000} collaboratively obtain the optimal
TS duration for 2-user IC, which is further extended to $K$-users
via introducing user-groups. In \cite{Timotheou2013} for multiple-input-single-output
(MISO) IC, PS ratio and power allocation is obtained to show that
maximal ratio transmission (MRT) based precoding outperforms zero-forcing
(ZF) in terms of EH. The work in \cite{6918468} is to improve harvested
energy and its consumption problem via power allocation, while keeping
fairness among users. In \cite{7037210,7054723}, antenna selection
is used for EH improvements. In \cite{6906324}, an upper bound on
EH is derived. In \cite{gupta2017improved,Zong2016430}, semi-definite
relaxation technique is leveraged to obtain suboptimal solutions via
convexifying the joint transceiver design problem. In \cite{7120020},
power splitting algorithm is proposed to maximize a linear-sum of
rate and energy objectives, where the coefficient of the linear-sum
decides the weight of these objectives.

\subsection{Contributions}

In this paper, a systematic precoding approach for SWIPT maximization
is investigated for the $K$-user MIMO-IC. From the above review,
it can be noted that in IA-SWIPT literature, authors have posed the
optimization problem as a linear sum of sum rate and harvested energy,
and sub-optimum solution have been computed convex relaxation tools
\cite{Zong2016430,gupta2017improved}. In this work, using chordal
distance (CD) decomposition, a systematic method is presented to obtain
the balanced precoding, which improves the trade-off between sum rate
and harvested energy. The proposed precoder, which is the key for
the rate-energy trade-off, can be obtainedvia maximizing the harvested
energy, or sum rate via tuning the value of chordal distance. EH analysis
shows the guaranteed improvement of energy for the proposed formulation.
Simulation results for different IA methods have been compared. These
results show that the IA methods utilizing direct channels, such as
MMSE and max-SINR algorithms, provide the better harvested power than
that of the IA methods, which does not utilize direct channels in
the precoder design including subspace method or leakage minimization
algorithm. Analog feedback automatically chooses the chordal distance,
which provides better EH and linear sum rate scaling at high SNR.
On the other hand, with quantized feedback, increasing the size of
codebook increases harvested energy, while suffering DoF losses. In
summary, the contribution of this paper can be listed as follows:

\subsubsection{Rate energy balanced precoding}

First, the maximum harvested energy achievable is obtained using the
precoders $\mathbf{V}^{EH}$, which defines the upper limit achievable
on EH. Then, we systematically derive the balanced precoding scheme
to improve SWIPT trade-off using CD decomposition. Beyond the upper
limit i.e., when the CD value between IA and the proposed precoder
is chosen greater than the CD value between the IA precoder and $\mathbf{V}^{EH}$,
the precoder $\mathbf{V}^{EH}$ provides the better sum rates. It
is also worth noting that the proposed method is much computationally
efficient, as compared to semi-definite programming. 

\subsubsection{Simple parameter design for constant rate loss}

We analyze the upper and lower bounds on the harvested energy. Tuning
the values of CD and PS ratio, the trade-off between sum rate and
harvested energy for IA networks can be observed. Further, the cases
of analog and quantized feedback are analyzed, where the CD value
is automatically set based on the feedback transmission power in the
analog feedback, and based on the size of codebook in the quantized
feedback. In both cases, harvested energy is shown to improve than
that in the case of IA precoder. Moreover, it is shown that it is
possible to obtain constant rate loss (i.e., linear sum rate scaling
with respect to SNR), while achieving the desired harvested energy
threshold.

\subsubsection{Which is the best IA scheme for harvesting?}

Simulation results verify the improvements and limits of balanced
precoder via the plots for rate-energy regions, and show that MSE
based IA schemes are better suited for SWIPT trade-offs than subspace
or leakage minimization schemes. Both analog and quantized feedback
improves the harvested energy, while achieving a constant rate loss
in the former case, and getting the rate loss proportional to the
chordal distance in the latter case.

\subsubsection*{Organization}

The interference channel model is given in section \ref{sec:System-Model}.
The next sections \ref{sec:Rate-Energy-Optimized-Precoding} and \ref{sec:Proposed-Balanced-precoding}
present energy optimized precoding and the proposed balanced precoding,
followed by the the cases of analog and quantized feedback in section
\ref{sec:Energy-Harvesting-withFB}. Simulation results are presented
in Section \ref{sec:Simulation-Results}. Section \ref{sec:Conclusion}
concludes the work.

\subsubsection*{Notations}

Scalars, vectors and matrices are represented by the lower case ($a$),
lower case bold face ($\mathbf{a}$) and upper case bold face ($\mathbf{A}$)
letters respectively. Conjugate, transpose, Hermitian transpose and
Kronecker product of matrices are denoted by $(\cdot)^{*}$, $(\cdot)^{T}$,
$(\cdot)^{\dagger}$ and $\otimes$ respectively. $\mathcal{CN}(\mathbf{\mu},\mathbf{R})$
represents a circularly symmetric complex Gaussian random vector with
mean $\mathbf{\mu}$ and covariance matrix $\mathbf{R}$. The notations
$\|\cdot\|$ and $\|\cdot\|_{F}$ denote the $l_{2}$ norm and Frobenious
norm. $\text{vec}(\mathbf{X})$ denotes the column-wise vector representation
of matrix $\mathbf{X}$. $\mathcal{D}(\mathbf{A}_{i})$ denotes a
block diagonal matrix with matrices $\mathbf{A}_{i}$ as its block
diagonal components. $\mathbb{O}(\mathbf{X})$ denotes the orthonormal
part of the QR decomposition of $\mathbf{X}$ \cite{Razavi2015}.
$\mathbf{X}$ is unitary means $\mathbf{X}\mathbf{X}^{\dagger}=\mathbf{X}^{\dagger}\mathbf{X}=\mathbf{I}$.
$\lambda_{\max}(\mathbf{A})$, $\nu_{\max}(\mathbf{A})$, $\nu_{1:d}(\mathbf{A})$
denote the maximum eigenvalue, the corresponding eigenvector of $\mathbf{A}$,
and the matrix with columns being the eigenvectors corresponding to
$d$-largest eigenvalues. $\delta_{ij}$ is Kronecker delta, which
takes value 1 when $i=j$ and $0$ otherwise. 

\section{\label{sec:System-Model}System Model}

\begin{figure}
\centering \includegraphics[width=8.9cm]{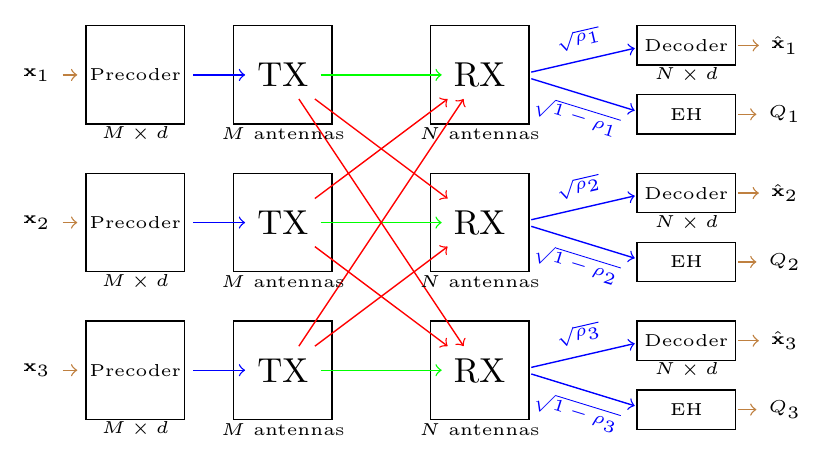}

\caption{$K=3$ user interference channel $\left(M\times N,d\right)^{K}$ with
energy harvesters. \label{fig:-user-interference}}

\end{figure}

Consider an IA-feasible MIMO interference channel \cite{Yetis2010,Gonzalez2014a,Wang2012a}
with $K$ users as shown in Figure \ref{fig:-user-interference}.
 Each user pair has $M$ transmit antennas, $N$ receive antennas
and $d$ independent data streams to be communicated. This system
is represented by the notation $(M\times N,d)^{K}$ \cite{Yetis2010}.
Let $\mathbf{x}_{k}$ of size $d\times1$ denote the transmit vector
of the $k^{th}$ user, distributed as $\mathcal{CN}(\mathbf{0},p_{k}\mathbf{I}_{d})$
with power $p_{k}=\frac{P_{k}}{d}$, where $P_{k}=\text{tr}\left(\mathbb{E}\left\{ \mathbf{x}_{k}\mathbf{x}_{k}^{\dagger}\right\} \right),\forall k$.
The MIMO channel matrix between the $j^{th}$ transmitter and $k^{th}$
receiver is denoted by $\mathbf{H}_{kj}\in\mathbb{C}^{N\times M}$.
The received signal at the $k^{th}$ user is given as 
\begin{equation}
\mathbf{y}_{k}=\mathbf{H}_{kk}\mathbf{V}_{k}\mathbf{x}_{k}+\sum_{j\neq k}\mathbf{H}_{kj}\mathbf{V}_{j}\mathbf{x}_{j}+\mathbf{n}_{k},\label{eq:yk1}
\end{equation}
where the precoder $\mathbf{V}_{k}=[\mathbf{v}_{k1},\,\mathbf{v}_{k2},\,\ldots,\,\mathbf{v}_{kd}]\in\mathbb{C}^{M\times d}$
is an orthonormal matrix employed at the transmitter and satisfies
the constraint $\mathbf{v}_{ki}^{\dagger}\mathbf{v}_{kj}=\delta_{ij},\,\forall k,\,i,\,j$.
The quantity $\mathbf{n}_{k}$ denotes zero mean additive white Gaussian
noise (AWGN), distributed as $\mathcal{CN}(\mathbf{0},\sigma^{2}\mathbf{I}_{N})$.
The first term in the above equation represents the desired signal
component, while the second and third terms correspond to the interference
and noise components respectively.

\subsection{Information Decoding }

Each receiver adaptively splits the received signal into two flows,
i.e., one part goes to the RF-EH circuits for energy storage, while
the other part is downconverted to for decoding the information. Let
$\rho_{k}\in[0,1]$ be the power splitting ratio, , which denotes
the portion of the received signal power assigned for ID, and the
remaining $(1-\rho_{k})$ portion allocated for harvesting energy.
Before ID, the split signal is further corrupted by additional circuit
noise $\mathbf{w}_{k}$, due to the non-ideal splitters, non-ideal
RF-to-baseband signal conversion, and thermal noise \cite{Zong2016430}.
Therefore, after splitting, the signal for ID can be expressed as
\begin{align}
\mathbf{y}_{k}^{ID} & =\sqrt{\rho_{k}}\mathbf{y}_{k}+\mathbf{w}_{k},\\
 & =\sqrt{\rho_{k}}\left(\mathbf{H}_{kk}\mathbf{V}_{k}\mathbf{x}_{k}+\sum_{j\neq k}\mathbf{H}_{kj}\mathbf{V}_{j}\mathbf{x}_{j}+\mathbf{n}_{k}\right)+\mathbf{w}_{k},
\end{align}
where $\mathbf{w}_{k}\sim\mathcal{CN}(\mathbf{0},\delta^{2}\mathbf{I}_{N})$
represents the power splitting circuit noise vector at receiver $k$.
The effective noise in the above equation can be written as $\mathbf{n}_{k}+\frac{\mathbf{w}_{k}}{\sqrt{\rho_{k}}}\sim\mathcal{CN}\left(\mathbf{0},\sigma_{ID}^{2}\mathbf{I}_{N}\right)$,
where $\sigma_{ID}^{2}=\sigma^{2}\left(1+\frac{\delta^{2}}{\rho_{k}\sigma^{2}}\right)$.
Further, the signal is processed via a linear receiver $\mathbf{U}_{k}$,
where $\mathbf{U}_{k}=[\mathbf{u}_{k1},\,\mathbf{u}_{k2},\,\ldots,\,\mathbf{u}_{kd}]\in\mathbb{C}^{N\times d}$
denotes an orthonormal decoding matrix, and can be obtained from minimizing
MSE \cite{Razavi2015,8379356}. 

\textcolor{blue}{\emph{Remark}}\textcolor{blue}{: For IA, it can be
assumed that the CSI values ($\mathbf{H}_{kj},\forall j$) are available
at the $k^{th}$ receiver. Receiver cooperation can be used to compute
IA precoders and decoders, i.e., one node collects different CSIs,
performs IA procedure, and then forwards the precoders/decoders to
the respective nodes. This process is more energy efficient and requires
a lower overhead \cite{Garg2017cqpq}. }

\subsubsection{IA feasibility}

In the above equation, to cancel the interference component and preserve
the desired signal, the precoders $\left\{ \mathbf{V}_{k},\forall k\right\} $
and the decoders $\left\{ \mathbf{U}_{k},\forall k\right\} $ should
be chosen to satisfy the following equations 
\begin{eqnarray}
\mathbf{U}_{k}^{\dagger}\mathbf{H}_{kj}\mathbf{V}_{j} & = & \mathbf{0},\,\forall j\neq k\forall k,\label{eq:matia1-1}\\
\text{rank}(\mathbf{U}_{k}^{\dagger}\mathbf{H}_{kk}\mathbf{V}_{k}) & = & d,\,\forall k.\label{eq:matia2-1}
\end{eqnarray}
In order to find possible solutions for $\left\{ \mathbf{V}_{k},\forall k\right\} $
and $\left\{ \mathbf{U}_{k},\forall k\right\} $, the system must
be IA-feasible. Characterizations of IA-feasible systems are given
in \cite{Yetis2010,Gonzalez2014a,Wang2012a}. \cite{Yetis2010}, and
others \cite{Bresler2014d,Bresler2014c,Razaviyayn2012b} have demonstrated
that feasible systems must necessarily be proper, which requires the
number of equations in \eqref{eq:matia1-1} to be lower than the number
of variables, i.e., $M+N-(K+1)d\geq0.$ In addition to the proper
condition, \cite{Gonzalez2014a} has shown that feasibility can be
verified by testing the surjectivity of the mapping proposed therein.
More specifically, \cite{Wang2012a} and \cite{Bresler2014d} have
shown that a proper system is feasible when either $M$ or $N$ is
divisible by $d$, or the system is symmetric, i.e., $M=N$. Further,
\cite{9097459} presents a uniqueness condition and ensures the IA-feasibility
and the global maximum sum rate (or minimum MSE). Thus, we utilize
the condition 
\begin{equation}
M+N-(K+2)d\geq0
\end{equation}
along with the IA algorithm from \cite{9097459} to get the IA-solution.

\subsubsection{Sum rate}

The resulting sum rate at the $k^{th}$ destination can be expressed
as
\begin{equation}
R_{k}=\log_{2}\left|\mathbf{I}_{d}+p_{k}\bar{\mathbf{H}}_{kk}\bar{\mathbf{H}}_{kk}^{\dagger}\left(\sigma_{ID}^{2}\mathbf{I}_{d}+\sum_{\forall j\neq k}p_{j}\bar{\mathbf{H}}_{kj}\bar{\mathbf{H}}_{kj}^{\dagger}\right)^{-1}\right|,\label{eq:rk}
\end{equation}
where $\bar{\mathbf{H}}_{kj}=\mathbf{U}_{k}^{\dagger}\mathbf{H}_{kj}\mathbf{V}_{j}$.
If the interference components are perfectly canceled i.e. $\bar{\mathbf{H}}_{kj}=\mathbf{0},\forall j\neq k$,
we have 
\begin{eqnarray}
R_{k,per} & = & \log_{2}\left|\mathbf{I}_{d}+\frac{P_{k}}{d\sigma_{ID}^{2}}\bar{\mathbf{H}}_{kk}\bar{\mathbf{H}}_{kk}^{\dagger}\right|\\
 & = & \sum_{i=1}^{d}\log_{2}\left(1+\frac{P_{k}}{d\sigma_{ID}^{2}}|\sigma_{kki}|^{2}\right),
\end{eqnarray}
where $\sigma_{kki}\forall i=1,\ldots,d$ are the singular values
of $\bar{\mathbf{H}}_{kk}$. 

\subsection{Harvested Energy}

The second part of the splitted received signal for energy storage
at receiver $k$ can be written as 
\begin{equation}
\mathbf{y}_{k}^{EH}=\sqrt{\bar{\rho}_{k}}\mathbf{y}_{k},\forall k,
\end{equation}
with $\bar{\rho}_{k}=1-\rho_{k}$. The corresponding average harvested
energy that can be stored at receiver $k$ can be expressed as 
\begin{align}
Q_{k} & =\zeta\mathbb{E}\left\{ \|\mathbf{y}_{k}^{EH}\|^{2}\right\} \\
 & \approx\zeta\bar{\rho}_{k}\sum_{j=1}^{K}\frac{P_{j}}{d}\|\mathbf{H}_{kj}\mathbf{V}_{j}\|_{F}^{2},\label{eq:max_EH-Qk}
\end{align}
where $0<\zeta<1$ represents the power conversion efficiency for
EH, which is assumed to be equal for all receivers in the paper. Note
that the noise power $\zeta(1-\rho_{k})N\sigma^{2}$ is negligible
and hence, is omitted in the above equation.

\section{\label{sec:Rate-Energy-Optimized-Precoding}Energy Optimized Precoding
Method}

In this section, we first derive the precoders that achieve the maximum
value of harvested energy, followed by the rate-loss analysis via
chordal distance. 

\subsection{\label{subsec:Maximum-EH-Precoding}Precoding for the maximum EH }

The maximization problem for the total harvested energy with respect
to precoders, subject to orthogonality constraints on the precoders
can be cast as 
\begin{subequations}
\begin{align}
\left\{ \mathbf{V}_{j}^{EH},\forall j\right\} =\arg\max_{\mathbf{V}_{j},\forall j} & \zeta\sum_{k}\bar{\rho}_{k}\sum_{j}\frac{P_{j}}{d}\|\mathbf{H}_{kj}\mathbf{V}_{j}\|_{F}^{2}\\
\text{subject to } & \|\mathbf{V}_{j}\|_{F}^{2}\leq d,\forall j.
\end{align}
\end{subequations}
For each precoder, one can decouple the above problem, and the solution
for the $j^{th}$ precoder $\mathbf{V}_{j}$ can be given using the
dominant eigenvectors corresponding to $d$ maximum eigenvalues of
the $\sum_{k}\bar{\rho}_{k}\mathbf{H}_{kj}^{\dagger}\mathbf{H}_{kj}$,
i.e., 
\begin{align}
\mathbf{V}_{j}^{EH} & =\arg\max_{\|\mathbf{V}_{j}\|_{F}^{2}\leq d}\text{tr}\left(\mathbf{V}_{j}^{\dagger}\mathbf{H}_{j}^{\dagger}\mathbf{H}_{j}\mathbf{V}_{j}\right)\label{eq:max_EH-Vj}\\
 & =\nu_{1:d}\left[\mathbf{H}_{j}^{\dagger}\mathbf{H}_{j}\right]=\mathbf{W}_{j}^{[1]},\nonumber 
\end{align}
where $\mathbf{H}_{j}^{T}=\left[\bar{\rho}_{1}\mathbf{H}_{1j}^{T},\ldots,\bar{\rho}_{K}\mathbf{H}_{Kj}^{T}\right]$
denotes a stack of the channel matrices, and $\mathbf{W}_{j}^{[1]}$
is computed via the eigenvalue decomposition (EVD) i.e., 
\begin{equation}
\mathbf{H}_{j}^{\dagger}\mathbf{H}_{j}=\sum_{k}\bar{\rho}_{k}\mathbf{H}_{kj}^{\dagger}\mathbf{H}_{kj}=\mathbf{W}_{j}\mathbf{\Lambda}_{j}\mathbf{W}_{j}^{\dagger},\label{eq:EVD_EH}
\end{equation}
with $\mathbf{W}_{j}=\left[\mathbf{W}_{j}^{[1]},\mathbf{W}_{j}^{[2]}\right]$
and $\mathbf{\Lambda}_{j}=\mathcal{D}\left(\lambda_{ji},i=1\ldots,M\right)$
such that $\lambda_{j1}\geq\ldots\geq\lambda_{jM}$ being in the descending
order. The matrices $\mathbf{W}_{j}^{[1]}$ and $\mathbf{W}_{j}^{[2]}$
are orthonormal matrices of size $M\times d$ and $M\times M-d$,
respectively. In the following, we define chordal distance in order
to analyze the rate loss. 

\subsection{Chordal Distance}
\begin{defn}
For two orthonormal matrices $\mathbf{V},\hat{\mathbf{V}}\in\mathbb{C}^{M\times d}$
such that $\hat{\mathbf{V}}^{\dagger}\hat{\mathbf{V}}=\mathbf{V}^{\dagger}\mathbf{V}=\mathbf{I}_{d}$,
the chordal distance between these matrices can be defined as 
\begin{eqnarray}
d_{c}^{2}(\hat{\mathbf{V}},\mathbf{V}) & = & \frac{1}{2}\|\mathbf{V}\mathbf{V}^{\dagger}-\hat{\mathbf{V}}\hat{\mathbf{V}}^{\dagger}\|_{F}^{2}=d-\|\mathbf{V}^{\dagger}\hat{\mathbf{V}}\|_{F}^{2}.\label{eq:chord_dist_mat-1}
\end{eqnarray}
\end{defn}
Note that the orthonormal matrices $\mathbf{V}$ and $\hat{\mathbf{V}}$
represent $d$ dimensional subspaces of $M$ dimensional space, i.e.,
$\mathbf{V}$ and $\hat{\mathbf{V}}$ lie on a Grassmannian manifold
$\mathcal{G}_{M,d}$, which is a collection of all such $d$ dimensional
subspaces. The chordal distance represents the distance between the
subspaces spanned by these matrices. Thus, two orthonormal matrices
who represent the same column space, will have zero distance. The
CD value between two unit-norm vectors (say $\mathbf{v}_{1},\mathbf{v}_{2}\in\mathcal{G}_{M,1}$),
is equivalent to computing the inner-product between them, i.e. $1-\left|\mathbf{v}_{1}^{\dagger}\mathbf{v}_{2}\right|^{2}$. 

Further, given two matrices in $\mathcal{G}_{M,d}$, one matrix can
be expressed into the other one via the CD decomposition from \cite[Lemma 1]{Ravindran2008}.
The following lemma states the modified CD decomposition, where the
modification comes from splitting the null space of dimension $M-d$
into a product of two variables.
\begin{lem}
\label{lem:MAT-CD-decomp-1}The two matrices $\hat{\mathbf{V}}$ and
$\mathbf{V}$ (such that $\hat{\mathbf{V}}^{\dagger}\hat{\mathbf{V}}=\mathbf{V}^{\dagger}\mathbf{V}=\mathbf{I}_{d}$)
admits the following decomposition \cite[Lemma 1]{9447191,Ravindran2008}
\begin{equation}
\mathbf{V}=\hat{\mathbf{V}}\mathbf{X}\mathbf{Y}+\hat{\mathbf{V}}^{\text{null}}\mathbf{S}\mathbf{Z},\label{eq:cd_decomp-1}
\end{equation}
where $\mathbf{V},\,\hat{\mathbf{V}}\in\mathbb{C}^{M\times d}$, $\hat{\mathbf{V}}_{j}^{\text{null}}=\text{null}(\hat{\mathbf{V}}_{j})\in\mathbb{C}^{M-d\times d},$
$\mathbf{X}\in\mathbb{C}^{d\times d}$ and $\mathbf{S}\in\mathbb{C}^{M-d\times d}$
are orthonormal matrices, $\mathbf{Y},\,\mathbf{Z}\in\mathbb{C}^{d\times d}$
are upper triangular matrices with positive diagonal elements satisfying
\begin{eqnarray}
\mathrm{tr}(\mathbf{Z}^{\dagger}\mathbf{Z}) & = & d_{c}^{2}(\mathbf{V},\hat{\mathbf{V}})\\
\mathbf{Y}^{\dagger}\mathbf{Y} & = & \mathbf{I}_{d}-\mathbf{Z}^{\dagger}\mathbf{Z},\label{eq:decomp-YY-1}
\end{eqnarray}
 Moreover, $\mathbf{X}$ and $\mathbf{Y}$ are distributed independent
of each other, as is the pair $\mathbf{S}$ and $\mathbf{Z}$.
\end{lem}
\begin{IEEEproof}
A short proof is included in Appendix\ref{sec:Proof-of-CD-1} \cite{Ravindran2008}.
\end{IEEEproof}
It can be noted that this decomposition requires $M\geq2d$, which
is the case in interference alignment, wherein at least $2d$ dimensions
are required i.e., at least $d$ dimensions for the desired signal
and the remaining for the interference. 
\begin{cor}
\label{cor:CDD1}For IC, if two sets of precoders have zero chordal
distances, then the resulting rate and the harvested energy are same. 
\end{cor}
\begin{IEEEproof}
For two precoders $\mathbf{V}_{j}$ and $\hat{\mathbf{V}}_{j}$ such
that $\mathbf{V}_{j}=\hat{\mathbf{V}}_{j}\mathbf{X}_{j}\mathbf{Y}_{j}$
with $\mathbf{X}_{j}\mathbf{X}_{j}^{\dagger}=\mathbf{Y}_{j}\mathbf{Y}_{j}^{\dagger}=\mathbf{I}_{d},\forall j$,
the product $\mathbf{V}_{j}\mathbf{V}_{j}^{\dagger}=\hat{\mathbf{V}}_{j}\hat{\mathbf{V}}_{j}^{\dagger},\forall j$
will be same, resulting in the same sum rate and the harvested energy. 
\end{IEEEproof}
In the above, it can be noted that two orthogonal matrices with zero
CD value will be termed as \emph{equivalent} matrices, rather than
considering the same matrix. 
\begin{cor}
\label{cor:Given-CD}Given an orthogonal matrix $\mathbf{V}$ and
the chordal distance value $z$ , the displacement precoder (with
respect to $\mathbf{V}$) via the CD decomposition, can be obtained
by relaxing the matrices $\mathbf{Y}$ and $\mathbf{Z}$ diagonal
matrices $\Sigma_{Y}$ and $\Sigma_{Z}$ such that $\Sigma_{Y}^{2}=\mathbf{I}_{d}-\Sigma_{Z}^{2}$,
that is, 
\begin{equation}
\mathbf{V}_{D}=\mathbf{V}\mathbf{X}\Sigma_{Y}+\mathbf{V}^{\text{null}}\mathbf{S}\Sigma_{Z}.
\end{equation}
\end{cor}

\begin{IEEEproof}
From Lemma \ref{lem:MAT-CD-decomp-1}, given the CD value $z$, the
desired displacement matrix with respect to $\mathbf{V}$ can be written
as $\mathbf{V}\bar{\mathbf{X}}\mathbf{Y}+\mathbf{V}^{\text{null}}\bar{\mathbf{S}}\mathbf{Z}$.
The chordal distance can then be equated as 
\begin{subequations}
\begin{align}
z & =d_{c}^{2}\left(\mathbf{V}\bar{\mathbf{X}}\mathbf{Y}+\mathbf{V}^{\text{null}}\bar{\mathbf{S}}\mathbf{Z},\mathbf{V}\right)\\
 & \stackrel{(a)}{=}d_{c}^{2}\left(\mathbf{V}\bar{\mathbf{X}}\mathbf{U}_{Y}\Sigma_{Y}\mathbf{V}_{Y}^{\dagger}+\mathbf{V}^{\text{null}}\bar{\mathbf{S}}\mathbf{U}_{Z}\Sigma_{Z}\mathbf{V}_{Y}^{\dagger},\mathbf{V}\right)\\
 & \stackrel{(b)}{=}d_{c}^{2}\left(\mathbf{V}\mathbf{X}\Sigma_{Y}+\mathbf{V}^{\text{null}}\mathbf{S}\Sigma_{Z},\mathbf{V}\mathbf{V}_{Y}\right)\\
 & \stackrel{(c)}{=}d_{c}^{2}\left(\mathbf{V}\mathbf{X}\Sigma_{Y}+\mathbf{V}^{\text{null}}\mathbf{S}\Sigma_{Z},\mathbf{V}\right)\\
 & =d_{c}^{2}\left(\mathbf{V}_{D},\mathbf{V}\right),
\end{align}
\end{subequations}
where in $(a)$, the SVD is used for $\mathbf{Z}=\mathbf{U}_{Z}\Sigma_{Z}\mathbf{V}_{Y}^{\dagger}$
and $\mathbf{Y}=\mathbf{U}_{Y}\Sigma_{Y}\mathbf{V}_{Y}^{\dagger}$
with the same right singular vectors owing to the constraint $\mathbf{Y}^{\dagger}\mathbf{Y}=\mathbf{I}_{d}-\mathbf{Z}^{\dagger}\mathbf{Z}$;
in $(b)$, $\mathbf{S}=\bar{\mathbf{S}}\mathbf{U}_{Z}$, $\mathbf{X}=\bar{\mathbf{X}}\mathbf{U}_{Y}$
are substituted; in $(c)$, the fact that the chordal distance is
unchanged for unitary multiplication $\mathbf{V}_{Y}$,  is used,
yielding that the matrices $\mathbf{Z}$ and $\mathbf{Y}$ can be
relaxed to diagonal ones. 
\end{IEEEproof}

\subsection{Rate loss upper bound for EH based precoding}

For the precoding in \eqref{eq:max_EH-Vj}, the resultant maximum
harvested energy can be given as the sum of the $d$ dominant eigenvalues
of $\sum_{k}\bar{\rho}_{k}\mathbf{H}_{kj}^{\dagger}\mathbf{H}_{kj}$.
Note that the precoding in \eqref{eq:max_EH-Vj} does not consider
the effect of interference on information decoding. However, the resulting
precoders may partially align the interference. This partial interference
alignment can be measured using the chordal distance value between
EH precoders and the ideal IA precoders as 
\begin{equation}
z_{j}^{EH}=d_{c}^{2}(\mathbf{V}_{j},\mathbf{V}_{j}^{EH}),\forall j,
\end{equation}
where $\mathbf{V}_{k},\,\forall k$ denote IA-precoders. The above
chordal distance represents the displacement of $\mathbf{V}_{j}^{EH}$
with respect to $\mathbf{V}_{j}$, and is independent of SNR values.
The more the distance, the more will be interference, and the less
will be the sum rates. Therefore, in the rate-energy trade-off, it
is essential to specify the allowable sum rate losses in the system,
which is characterized in the following result \cite{8379356}.
\begin{lem}
\label{lem:Rate-loss-basic}(Rate Loss Upper Bound (RLUB)) For an
IC $(M\times N,d)^{K}$, employing an imperfect precoder at the source
instead of IA precoder results in the rate loss $\Delta R_{k}$, the
expected value of which can be bounded at the $k^{th}$ receiver as
\begin{align}
\mathbb{E}\left\{ \Delta R_{k}\right\}  & <d\log_{2}\left(1+\frac{P}{\sigma_{ID}^{2}}M_{d}\sum_{j\neq k}z_{j}\right),\label{eq:lemmarateloss}
\end{align}
with $M_{d}=\frac{M}{d(M-d)}$, and $z_{k}=\mathbb{E}d_{c}^{2}(\mathbf{V}_{k},\hat{\mathbf{V}}_{k})$
being the average chordal distance between the imperfect precoder
and the IA precoder.
\end{lem}
\begin{IEEEproof}
Proof is given in Appendix\ref{subsec:Proof-of-RLUB} \cite[Lemma 4]{8379356}.
\end{IEEEproof}

\subsection{Problem Formulation}

In literature for the SWIPT precoding \cite{Krikidis2014104,Zhao2015},
an optimization problem is formulated, where a linear sum of the sum
rate and sum harvested energy is maximized subjected to precoder constraints
and quality-of-service constraints as 
\begin{subequations}
\begin{gather}
\max_{\mathbf{V}_{j},\forall j}\sum_{k}R_{k}\left(\mathbf{V}_{j},\forall j\big|\mathbf{H}_{j}\right)+\nu Q_{k}\left(\mathbf{V}_{j},\forall j\big|\mathbf{H}_{j}\right)\\
\text{subject to }\|\mathbf{V}_{j}\|_{F}^{2}\leq d,\forall j,\\
\left|R_{j}-\bar{R}_{j}\right|\leq d\log_{2}c,\forall j,
\end{gather}
\end{subequations}
where $\nu$ is the weight controlling the preferred objective; and
$\bar{R}_{j}$ and $d\log_{2}c$ are the rate and the rate-loss constraint. 

Two objectives in the above are opposite in nature, i.e. sum rate
maximization leads to reduced harvested energy, and the maximization
of harvested energy degrades the sum rates. For the balanced precoding,
we prioritize the sum rate maximized precoder i.e. IA precoder $\mathbf{V}_{j}$,
and distort this precoder in such that the modified balanced precoder
satisfies the required constraints. In general, if the IA precoder
is degraded, it can result in severe rate losses, causing DoF loss.
Thus, to displace the IA precoder in a systematic way, we utilize
the chordal distance decomposition, in which the value of chordal
distance decides the degradation of precoders, that is, DoF losses.
From Lemma \ref{lem:Rate-loss-basic}, it can be seen that if the
distance is chosen proportional to the inverse of SNR, only a constant
rate loss is present,  that is, there is no-loss of DoFs. This value
of constant loss can be varied using the splitting ratio. 

For example, to keep the RLUB  a constant (say $d\log_{2}c$), the
required CD value can be given as 
\begin{equation}
z_{k}\leq\bar{z}(P)=\frac{c-1}{M_{d}(K-1)}\left(\frac{P}{\sigma_{ID}^{2}}\right)^{-1},
\end{equation}
which along with EH constraint can be written as 
\begin{equation}
z_{k}\leq\min\left(\bar{z}(P),z_{k}^{EH}\right).
\end{equation}
In the above, one should have $\bar{z}(P)\leq z_{k}^{EH}$; otherwise,
the sum rates would be worse due to the absence of IA, and in that
case, the EH-maximizing precoder will be the better choice. In high
SNR regime with analog feedback, these conditions can met easily,
since the CD value is inversely proportional to SNR. In low and mid-SNR
range, the value of splitting ratio $\rho$ can be finely tuned to
get the CD value $z_{k}$ within the limit. Thus, given IA precoders
and CD values with a specified constant RLUB, the balanced precoders
are computed and analyzed in the following sections.

\section{Proposed Balanced precoding Method\label{sec:Proposed-Balanced-precoding}}

Given the CD value $\left\{ z_{j},\forall j\right\} $ and the IA
precoders $\left\{ \mathbf{V}_{j},\forall j\right\} $, the objective
of precoder optimization reduces to the maximization of the harvested
energy, since the resulting sum rate with a given CD value provides
fixed rate loss. Then using Corollary \ref{cor:Given-CD}, the $j^{th}$
balanced precoder can be given as 
\begin{equation}
\mathbf{V}_{j}^{BAL}=\mathbf{V}_{j}\mathbf{X}_{j}\mathbf{Y}_{j}+\mathbf{V}_{j}^{\text{null}}\mathbf{S}_{j}\mathbf{Z}_{j},
\end{equation}
where the matrices $\mathbf{Y}_{j}$ and $\mathbf{Z}_{j}$ are diagonal;
$\mathbf{V}_{j}^{\text{null}}$ represents the left null space of
$\mathbf{V}_{j}$, i.e., $\mathbf{V}_{j}^{\text{null}}=\text{null}(\mathbf{V}_{j})\in\mathcal{G}_{M,M-d}$
and $\mathbf{V}_{j}^{\text{null}\dagger}\mathbf{V}_{j}=\mathbf{0}$. 

The maximization problem for the total harvested energy with respect
to finding the balanced precoding can be cast as 
\begin{subequations}
\begin{align}
\max_{\mathbf{S}_{j},\mathbf{Z}_{j},\mathbf{X}_{j},\mathbf{Y}_{j},\forall j}\sum_{k}\zeta\bar{\rho}_{k}\sum_{j}\frac{P_{j}}{d}\|\mathbf{H}_{kj}\mathbf{V}_{j}^{BAL}\|_{F}^{2}\label{eq:porb_state1}\\
\text{subject to }\mathbf{V}_{j}^{BAL}=\mathbf{V}_{j}\mathbf{X}_{j}\mathbf{Y}_{j}+\mathbf{V}_{j}^{\text{null}}\mathbf{S}_{j}\mathbf{Z}_{j},\forall j,\\
\text{tr}\left(\mathbf{Z}_{j}\mathbf{Z}^{\dagger}\right)=\text{tr}\left(\mathbf{I}-\mathbf{Y}_{j}\mathbf{Y}_{j}^{\dagger}\right)\leq z_{j},\forall j,\label{eq:z_j_con}\\
\mathbf{Z}_{j},\mathbf{Y}_{j}\text{ are diagonal matrices},\forall j,\\
\mathbf{X}_{j}^{\dagger}\mathbf{X}_{j}=\mathbf{X}_{j}\mathbf{X}_{j}^{\dagger}=\mathbf{I},\forall j,\\
\mathbf{S}_{j}^{\dagger}\mathbf{S}_{j}=\mathbf{I},\forall j.
\end{align}
\end{subequations}
It can be observed that one can decouple the above problem for each
$j^{th}$ precoder as 
\begin{subequations}
\begin{align}
\max_{\mathbf{S}_{j},\mathbf{Z}_{j},\mathbf{X}_{j},\mathbf{Y}_{j}}\left\Vert \mathbf{H}_{j}\mathbf{V}_{j}^{BAL}\right\Vert _{F}^{2}\label{eq:porb_state2}\\
\text{subject to }\mathbf{V}_{j}^{BAL}=\mathbf{V}_{j}\mathbf{X}_{j}\mathbf{Y}_{j}+\mathbf{V}_{j}^{\text{null}}\mathbf{S}_{j}\mathbf{Z}_{j},\label{eq:VBALCON}\\
\text{tr}\left(\mathbf{Z}_{j}\mathbf{Z}^{\dagger}\right)=\text{tr}\left(\mathbf{I}-\mathbf{Y}_{j}\mathbf{Y}_{j}^{\dagger}\right)\leq z_{j},\\
\mathbf{Z}_{j},\mathbf{Y}_{j}\text{ are diagonal matrices},\\
\mathbf{X}_{j}^{\dagger}\mathbf{X}_{j}=\mathbf{X}_{j}\mathbf{X}_{j}^{\dagger}=\mathbf{I},\label{eq:Xj_con}\\
\mathbf{S}_{j}^{\dagger}\mathbf{S}_{j}=\mathbf{I},
\end{align}
\end{subequations}
, whose solution is computed as follows. We first obtain $\mathbf{S}_{j}$,
followed by the computation of $\mathbf{Z}_{j}$ and $\mathbf{X}_{j}$,
which can be derived using either iterative or non-iterative method
given below. 

\subsection{Getting $\mathbf{S}_{j}$ }

The objective function using the triangle inequality in \eqref{eq:porb_state2}
can be bounded as 
\begin{align}
 & \left\Vert \mathbf{H}_{j}\left(\mathbf{V}_{j}\mathbf{X}_{j}\mathbf{Y}_{j}+\mathbf{V}_{j}^{\text{null}}\mathbf{S}_{j}\mathbf{Z}_{j}\right)\right\Vert _{F}\nonumber \\
 & \leq\left\Vert \mathbf{H}_{j}\mathbf{V}_{j}\mathbf{X}_{j}\mathbf{Y}_{j}\right\Vert _{F}+\left\Vert \mathbf{H}_{j}\mathbf{V}_{j}^{\text{null}}\mathbf{S}_{j}\mathbf{Z}_{j}\right\Vert _{F},\label{eq:HVABL_tri}
\end{align}
where the equality occurs when both $\mathbf{H}_{j}\mathbf{V}_{j}\mathbf{X}_{j}\mathbf{Y}_{j}$
and $\mathbf{H}_{j}\mathbf{V}_{j}^{\text{null}}\mathbf{S}_{j}\mathbf{Z}_{j}$
have same directions or are proportional to each other. It can be
noted that since both the matrix $\mathbf{V}_{j}$ and its null space
$\mathbf{V}_{j}^{\text{null}}$ are present in the above expression,
the equality cannot be achieved when $z_{j}>0$ or $\mathbf{Z}_{j}\neq\mathbf{0}$.
Best efforts can be performed in order to align these two matrices
via the following CD minimization problem as 
\begin{subequations}
\begin{flalign}
\min_{\mathbf{S}_{j},\mathbf{Z}_{j},\mathbf{X}_{j},\mathbf{Y}_{j}} & d_{c}^{2}\left(\mathbb{O}\left(\mathbf{H}_{j}\mathbf{V}_{j}\mathbf{X}_{j}\mathbf{Y}_{j}\right),\mathbb{O}\left(\mathbf{H}_{j}\mathbf{V}_{j}^{\text{null}}\mathbf{S}_{j}\mathbf{Z}_{j}\right)\right),\\
\stackrel{(a)}{=} & \min_{\mathbf{S}_{j}}d_{c}^{2}\left(\mathbb{O}\left(\mathbf{H}_{j}\mathbf{V}_{j}\right),\mathbb{O}\left(\mathbf{H}_{j}\mathbf{V}_{j}^{\text{null}}\mathbf{S}_{j}\right)\right),\\
\stackrel{(b)}{=} & \max_{\mathbf{S}_{j}^{\dagger}\mathbf{S}_{j}=\mathbf{I}}\text{tr}\left[\mathbf{D}_{Vj}\mathbf{V}_{j}^{\dagger}\mathbf{H}_{j}^{\dagger}\mathbf{H}_{j}\mathbf{V}_{j}^{\text{null}}\mathbf{S}_{j}\mathbf{D}_{Vnj}\right],
\end{flalign}
\end{subequations}
where in $(a)$, a property of orthogonalization is utilized; in \textbf{$(b)$},
the definition of CD, $\mathbb{O}\left(\mathbf{A}\right)=\mathbf{A}\left(\mathbf{A}^{\dagger}\mathbf{A}\right)^{-1/2}$,
$\mathbf{D}_{Vj}=\left(\mathbf{V}_{j}^{\dagger}\mathbf{H}_{j}^{\dagger}\mathbf{H}_{j}\mathbf{V}_{j}\right)^{-1/2}$,
and $\mathbf{D}_{Vnj}=\left(\mathbf{S}_{j}^{\dagger}\mathbf{V}_{j}^{\text{null}\dagger}\mathbf{H}_{j}^{\dagger}\mathbf{H}_{j}\mathbf{V}_{j}^{\text{null}}\mathbf{S}_{j}\right)^{-1/2}$
are substituted. From $(b)$, the solution can be computed by selecting
the columns in the same directions as $\mathbf{V}_{j}^{\text{null}\dagger}\mathbf{H}_{j}^{\dagger}\mathbf{H}_{j}\mathbf{V}_{j}$ as
\begin{align}
\mathbf{S}_{j} & =\mathbb{O}\left(\mathbf{V}_{j}^{\text{null}\dagger}\mathbf{H}_{j}^{\dagger}\mathbf{H}_{j}\mathbf{V}_{j}\mathbf{D}_{Vj}\mathbf{D}_{Vnj}\right)\\
 & \equiv\mathbb{O}\left(\mathbf{V}_{j}^{\text{null}\dagger}\mathbf{H}_{j}^{\dagger}\mathbf{H}_{j}\mathbf{V}_{j}\right),
\end{align}
where the equivalence can be considered due to the fact that $\mathbf{X}_{j}$,
$\mathbf{Y}_{j}$ and $\mathbf{Z}_{j}$ are unknown (or yet to be
designed based on $\mathbf{S}_{j}$), and thus, $\mathbf{S}_{j}$
can be independently computed first. Ltting $\mathbf{A}_{j}=\mathbf{V}_{j}^{\text{null}\dagger}\mathbf{H}_{j}^{\dagger}\mathbf{H}_{j}\mathbf{V}_{j}$,
the cross-term of two matrices reduces to 
\begin{align*}
 & \text{tr}\left(\mathbf{Y}_{j}^{\dagger}\mathbf{X}_{j}^{\dagger}\mathbf{V}_{j}^{\dagger}\mathbf{H}_{j}^{\dagger}\mathbf{H}_{j}\mathbf{V}_{j}^{\text{null}}\mathbf{S}_{j}\mathbf{Z}_{j}\right)=\text{tr}\left(\mathbf{Z}_{j}\mathbf{Y}_{j}^{\dagger}\mathbf{X}_{j}^{\dagger}\left(\mathbf{A}_{j}^{\dagger}\mathbf{A}_{j}\right)^{1/2}\right).
\end{align*}

\subsection{Getting $\mathbf{Z}_{j}$ and $\mathbf{X}_{j}$: an iterative approach}

From \eqref{eq:HVABL_tri}, squaring the terms on both sides yields
the Cauchy Schwarz's inequality 
\begin{subequations}
\begin{align}
 & \Re\text{tr}\left(\mathbf{Y}_{j}^{\dagger}\mathbf{X}_{j}^{\dagger}\mathbf{V}_{j}^{\dagger}\mathbf{H}_{j}^{\dagger}\mathbf{H}_{j}\mathbf{V}_{j}^{\text{null}}\mathbf{S}_{j}\mathbf{Z}_{j}\right)\label{eq:HVAL_cauchy}\\
 & \leq\left\Vert \mathbf{H}_{j}\mathbf{V}_{j}\mathbf{X}_{j}\mathbf{Y}_{j}\right\Vert _{F}\left\Vert \mathbf{H}_{j}\mathbf{V}_{j}^{\text{null}}\mathbf{S}_{j}\mathbf{Z}_{j}\right\Vert _{F},
\end{align}
\end{subequations}
which suggests that equivalently, the above cross-term can be maximized
to get the maximum harvested energy. 

Since the matrices $\mathbf{Y}_{j}$ and $\mathbf{Z}_{j}$ are relaxed
to be diagonal, the matrix $\mathbf{Y}_{j}=\mathcal{D}\left(y_{j1},\ldots y_{jd}\right)$
can be obtained from $\mathbf{Z}_{j}=\mathcal{D}\left(z_{j1},\ldots z_{jd}\right)$
using the constraint in \eqref{eq:z_j_con} and \eqref{eq:decomp-YY-1}
as 
\begin{equation}
y_{ji}=+\sqrt{1-z_{ji}^{2}},\forall i=1,\ldots,d,\label{eq:Y_chol1}
\end{equation}
satisfying the constraint in \eqref{eq:z_j_con}. The remaining components
of the CD decomposition can be computed as the solution of the following
optimization problem as 
\begin{align*}
 & \max_{\mathbf{Z}_{j},\mathbf{X}_{j}}\Re\text{tr}\left(\mathbf{Y}_{j}^{\dagger}\mathbf{X}_{j}^{\dagger}\mathbf{V}_{j}^{\dagger}\mathbf{H}_{j}^{\dagger}\mathbf{H}_{j}\mathbf{V}_{j}^{\text{null}}\mathbf{S}_{j}\mathbf{Z}_{j}\right),
\end{align*}
which is a non-convex problem due to the product of $\mathbf{Z}_{j}$
and $\mathbf{X}_{j}$. The efficient way to solve the problem is via
an iterative method, where $\mathbf{X}_{j}$ and $\mathbf{Z}_{j}$
are solved alternately. 

Given $\mathbf{Z}_{j}$ and $\mathbf{Y}_{j}$, the optimization problem
above can be reduced to a convex problem as 
\begin{subequations}
\begin{align}
 & \max_{\mathbf{X}_{j}}\Re\text{tr}\left(\mathbf{Y}_{j}^{\dagger}\mathbf{X}_{j}^{\dagger}\mathbf{V}_{j}^{\dagger}\mathbf{H}_{j}^{\dagger}\mathbf{H}_{j}\mathbf{V}_{j}^{\text{null}}\mathbf{S}_{j}\mathbf{Z}_{j}\right)\label{eq:optX}\\
 & \text{subject to }\|\mathbf{X}_{j}\|\leq1,
\end{align}
\end{subequations}
where the spectral norm constraint above leads to the same constraint
in \eqref{eq:Xj_con}. The solution for $\mathbf{X}_{j}$ can be obtaining
by choosing the same column directions as of $\mathbf{V}_{j}^{\dagger}\mathbf{H}_{j}^{\dagger}\mathbf{H}_{j}\mathbf{V}_{j}^{\text{null}}\mathbf{S}_{j}\mathbf{Z}_{j}\mathbf{Y}_{j}^{\dagger}$,
i.e., 
\begin{align}
\mathbf{X}_{j} & =\left[\frac{\mathbf{b}_{1}}{\|\mathbf{b}_{1}\|_{2}},\ldots,\frac{\mathbf{b}_{d}}{\|\mathbf{b}_{d}\|_{2}}\right]=\mathbf{B}_{j}\mathbf{D}_{Bj}^{-1},
\end{align}
where $\mathbf{B}_{j}=\mathbf{V}_{j}^{\dagger}\mathbf{H}_{j}^{\dagger}\mathbf{H}_{j}\mathbf{V}_{j}^{\text{null}}\mathbf{S}_{j}\mathbf{Z}_{j}\mathbf{Y}_{j}^{\dagger}=\left[\mathbf{b}_{1},\ldots,\mathbf{b}_{d}\right]$
and $\mathbf{D}_{Bj}=\mathcal{D}\left(\|\mathbf{b}_{1}\|_{2},\ldots,\|\mathbf{b}_{d}\|_{2}\right)\succeq\mathbf{0}$.
Note that the above $\mathbf{X}_{j}$ cannot be equivalently set to
$\mathbb{O}\left(\mathbf{V}_{j}^{\dagger}\mathbf{H}_{j}^{\dagger}\mathbf{H}_{j}\mathbf{V}_{j}^{\text{null}}\mathbf{S}_{j}\right)$,
since the above particular directions are important. Further, substituting
$\mathbf{X}_{j}$ in the trace yields the following result. 
\begin{prop}
\label{prop:trace-val-noneg}With the above selection of $\mathbf{X}_{j}$,
the trace-value is non-negative 
\[
\text{tr}\left(\mathbf{Y}_{j}^{\dagger}\mathbf{X}_{j}^{\dagger}\mathbf{V}_{j}^{\dagger}\mathbf{H}_{j}^{\dagger}\mathbf{H}_{j}\mathbf{V}_{j}^{\text{null}}\mathbf{S}_{j}\mathbf{Z}_{j}\right)=\text{tr}\left(\mathbf{B}_{j}^{\dagger}\mathbf{D}_{Bj}^{-1}\mathbf{B}_{j}\right)\geq0,
\]
where  the equality occurs when $z_{j}=0$.
\end{prop}
Next, given $\mathbf{Y}_{j}$, $\mathbf{X}_{j}$ and $z_{j}<z_{j}^{EH}$,
the diagonal matrix $\mathbf{Z}_{j}$ can be updated as 
\begin{subequations}
\begin{align}
 & \max_{\mathbf{Z}_{j}}\Re\text{tr}\left(\mathbf{Y}_{j}^{\dagger}\mathbf{X}_{j}^{\dagger}\mathbf{V}_{j}^{\dagger}\mathbf{H}_{j}^{\dagger}\mathbf{H}_{j}\mathbf{V}_{j}^{\text{null}}\mathbf{S}_{j}\mathbf{Z}_{j}\right)\label{eq:optZ}\\
 & \text{subject to }\|\mathbf{Z}_{j}\|_{F}\leq\sqrt{z_{j}},\label{eq:Z_con1}\\
 & \mathbf{Z}_{j}\text{ is a diagonal matrix},\label{eq:Z_con2diag}\\
 & \mathbf{0}\preceq\mathbf{Z}_{j}\preceq\mathbf{I},\label{eq:Z01con}
\end{align}
\end{subequations}
which is also a convex problem. We can equivalently recast the problem
for $\mathbf{z}_{j}^{T}=\left[z_{j1},\ldots z_{jd}\right]$ as 
\begin{subequations}
\begin{align}
 & \max_{\mathbf{z}_{j}}\mathbf{c}_{j}^{T}\mathbf{z}_{j}\\
 & \text{subject to }\|\mathbf{z}_{j}\|_{2}\leq\sqrt{z_{j}},\\
 & 0\leq z_{j,i}\leq1,\forall i=1,\ldots,d,
\end{align}
\end{subequations}
where the vector $\mathbf{c}_{j}=\left[c_{j1},\ldots,c_{jd}\right]$
and $c_{ji}=\left[\mathbf{Y}_{j}^{\dagger}\mathbf{X}_{j}^{\dagger}\mathbf{V}_{j}^{\dagger}\mathbf{H}_{j}^{\dagger}\mathbf{H}_{j}\mathbf{V}_{j}^{\text{null}}\mathbf{S}_{j}\right]_{i,i},\forall i=1,\ldots,d$.
The values $c_{ji},\forall i$ are real and non-negative from the
proposition \ref{prop:trace-val-noneg}. The solution of the above
problem is given by choosing $\mathbf{z}_{j}$ equal to $\mathbf{c}_{j}$
and scaling it to satisfy the norm constraint. Thus, we write  $z_{ji}=\min\left(\sqrt{z_{j}}\frac{c_{ji}}{\|\mathbf{c}_{j}\|},1\right)$,
and normalize the resulting entries to satisfy $\sum_{i\in\mathcal{I}}z_{ji}^{2}=z_{j}-\left(d-\left|\mathcal{I}\right|\right)$,
where $\mathcal{I}=\left\{ i:z_{ji}<1\right\} $, i.e., $z_{ji}\leftarrow\frac{z_{ji}}{\sum_{i\in\mathcal{I}}z_{ji}^{2}}\sqrt{z_{j}-\left(d-\left|\mathcal{I}\right|\right)},\forall i\in\mathcal{I}$.

\subsection{Algorithm}

\begin{algorithm}
\begin{algorithmic}[1]

\Require{$\mathbf{H}_{j}$, $\mathbf{V}_{j}$ and $z_{j}$.}

\Ensure{$\mathbf{V}_{j}^{BAL}$.}

\If{ $z_{j}>z_{j}^{EH}$}

\State{Return $\mathbf{V}_{j}^{BAL}=\mathbf{V}_{j}^{EH}$.}

\Else

\State{Compute $\mathbf{S}_{j}=\mathbb{O}\left(\mathbf{V}_{j}^{\text{null}\dagger}\mathbf{H}_{j}^{\dagger}\mathbf{H}_{j}\mathbf{V}_{j}\right)$.}

\State{Initialize $\mathbf{Z}_{j}=\sqrt{\frac{z_{j}}{d}}\mathbf{I}$
and $\mathbf{Y}_{j}$ by \eqref{eq:Y_chol1}.}

\State{\label{step}Solve \eqref{eq:optX} to get $\mathbf{X}_{j}$.}

\State{Solve \eqref{eq:optZ} to get $\mathbf{Z}_{j}$.}

\State{Get $\mathbf{Y}_{j}$ by \eqref{eq:Y_chol1}.}

\State{Go to step \ref{step} until convergence. }

\State{Return $\mathbf{V}_{j}^{BAL}$ via \eqref{eq:VBALCON}.}

\EndIf

\end{algorithmic}\caption{Iterative CD decomposition procedure.\label{alg:Iterative-CD-decomposition}}
\end{algorithm}

\begin{figure}
\centering \includegraphics[width=8.9cm]{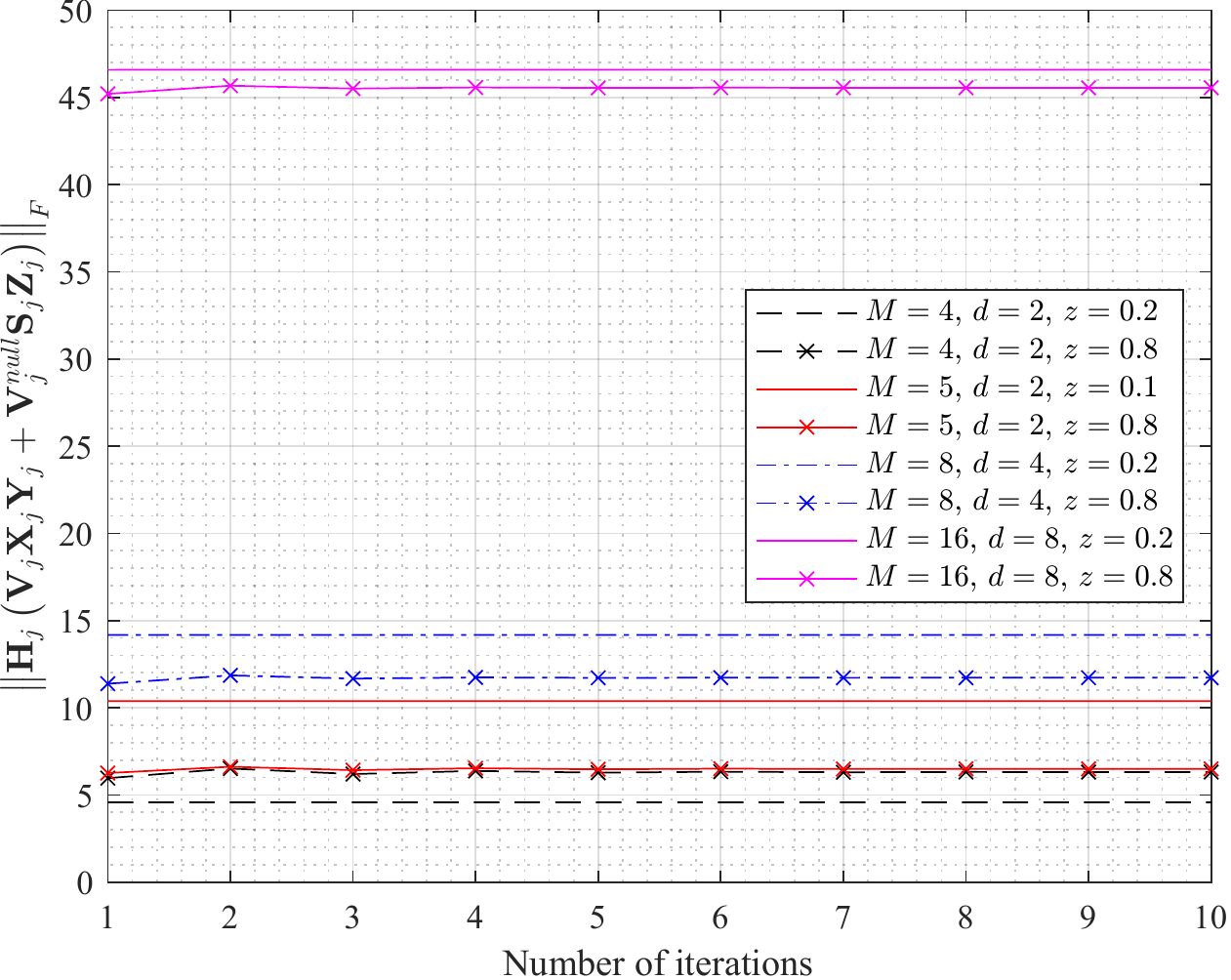}

\caption{Norm value versus the number of iterations for different interference
channels and different values of chordal distance. \label{fig:Norm-value-versus}}
\end{figure}

Now, with all components obtained, the resulting balanced precoder
can be computed via \eqref{eq:VBALCON}. The summary of this procedure
is given in Algorithm \ref{alg:Iterative-CD-decomposition}. If $z_{j}>z_{j}^{EH}$,
we choose energy optimized precoder as the balanced precoder $\mathbf{V}_{j}^{BAL}=\mathbf{V}_{j}^{EH}$.
Regarding the convergence, it can be seen that since both $\mathbf{Z}_{j}$
and $\mathbf{X}_{j}$ maximize the same linear objective, thus convergence
is guaranteed with a global optimum value, as plotted in Figure \ref{fig:Norm-value-versus}.
 Regarding the number of iterations, we observe via simulations that
it takes only a few ($4\text{ to }8$) iterations to converge. 

\subsection{Getting $\mathbf{Z}_{j}$ and $\mathbf{X}_{j}$: a non-iterative
approach}

Here, we present a suboptimal non-iterative method to compute $\mathbf{X}_{j}$
and $\mathbf{Z}_{j}$. This method is based on upper bound in the
equation \eqref{eq:HVABL_tri}. In \eqref{eq:HVABL_tri}, applying
the max-operator on both sides yields 
\begin{align}
 & \max_{\mathbf{S}_{j},\mathbf{Z}_{j},\mathbf{X}_{j},\mathbf{Y}_{j}}\left\Vert \mathbf{H}_{j}\left(\mathbf{V}_{j}\mathbf{X}_{j}\mathbf{Y}_{j}+\mathbf{V}_{j}^{\text{null}}\mathbf{S}_{j}\mathbf{Z}_{j}\right)\right\Vert _{F}\nonumber \\
 & \leq\max_{\mathbf{X}_{j},\mathbf{Y}_{j}}\left\Vert \mathbf{H}_{j}\mathbf{V}_{j}\mathbf{X}_{j}\mathbf{Y}_{j}\right\Vert _{F}+\max_{\mathbf{S}_{j},\mathbf{Z}_{j}}\left\Vert \mathbf{H}_{j}\mathbf{V}_{j}^{\text{null}}\mathbf{S}_{j}\mathbf{Z}_{j}\right\Vert _{F},
\end{align}
Thus, for a lower complexity solution, we solve the right hand side
get the components for the balanced precoding. 

Given $z_{j}<z_{j}^{EH}$, the matrix $\mathbf{Z}_{j}$ can be obtained
to maximize the harvested power as 
\begin{align}
\mathbf{Z}_{j}= & \arg\max_{\mathbf{Z}_{j}\forall j}\|\mathbf{H}_{j}\mathbf{V}_{j}^{\text{null}}\mathbf{S}_{j}\mathbf{Z}_{j}\|_{F}^{2}\label{eq:Z1}\\
 & \text{subject to \eqref{eq:Z_con1}, \eqref{eq:Z_con2diag}, \eqref{eq:Z01con}}.\nonumber 
\end{align}
The above problem can be simplified as 
\begin{subequations}
\begin{align}
 & \max_{0\leq z_{ji}\leq1,\forall i}\sum_{i}z_{ji}^{2}f_{ji}\\
 & \text{subject to }\sum_{i}z_{ji}^{2}\leq z_{j},
\end{align}
\end{subequations}
where the values $f_{ji}=\left[\mathbf{S}_{j}^{\dagger}\mathbf{V}_{j}^{\text{null}\dagger}\mathbf{H}_{j}^{\dagger}\mathbf{H}_{j}\mathbf{V}_{j}^{\text{null}}\mathbf{S}_{j}\right]_{i,i},\forall i=1,\ldots,d$
are real and non-negative. For the solution of the above optimization,
we write 
\begin{equation}
z_{ji}^{2}=\min\left(z_{j}\frac{f_{ji}}{\sum_{i}f_{ji}},1\right),\forall i=1,\ldots,d,
\end{equation}
and normalize the resulting entries ($\mathcal{I}=\left\{ i:z_{ji}<1\right\} $)
to satisfy $\sum_{i\in\mathcal{I}}z_{ji}^{2}=z_{j}-\left(d-\left|\mathcal{I}\right|\right)$.

Next, the matrix $\mathbf{Y}_{j}$ can be computed using \eqref{eq:Y_chol1}.
Further, the matrix $\mathbf{X}_{j}$ can be chosen as 
\begin{align}
\mathbf{X}_{j}= & \arg\max_{\|\mathbf{X}_{j}\|\leq1}\|\mathbf{H}_{j}\mathbf{V}_{j}\mathbf{X}_{j}\mathbf{Y}_{j}\|_{F}^{2}\label{eq:XjOPT}\\
= & \bar{\mathbf{B}}_{j}\bar{\mathbf{D}}_{Bj}^{-1},\label{eq:X_QR}
\end{align}
where $\bar{\mathbf{B}}_{j}=\mathbf{V}_{j}^{\dagger}\mathbf{H}_{j}^{\dagger}\mathbf{H}_{j}\mathbf{V}_{j}\mathbf{Y}_{j}^{-1}=\left[\bar{\mathbf{b}}_{1},\ldots,\bar{\mathbf{b}}_{d}\right]$
and $\bar{\mathbf{D}}_{Bj}=\mathcal{D}\left(\|\bar{\mathbf{b}}_{1}\|_{2},\ldots,\|\bar{\mathbf{b}}_{d}\|_{2}\right)$.
The resulting $j^{th}$ precoder $\mathbf{V}_{j}^{BAL}$ can be given
via \eqref{eq:VBALCON}. 

\subsection{Computational complexity}

The product $\mathbf{H}_{j}^{\dagger}\mathbf{H}_{j}$ need $\mathcal{O}\left(M^{2}NK\right)$
operations. For $\mathbf{S}_{j}$, the product and $\mathbb{O}(\cdot)$
need $\mathcal{O}\left(NK\cdot M(M-d)+NKd\cdot M+NK(M-d)\cdot M\right)=\mathcal{O}\left(NKM^{2}\right)$
and $\mathcal{O}\left(d^{2}\cdot(M-d)+d^{3}\right)=\mathcal{O}\left(Md^{2}\right)$,
respectively. The rest of operations are below $\mathcal{O}\left(Md^{3}\right)$
or \emph{$\mathcal{O}\left(M^{3}\right)$}. Thus, if $M=N$, Algorithm
\ref{alg:Iterative-CD-decomposition} has $\mathcal{O}\left(M^{3}K+Md^{2}N_{I}\right)\approx\mathcal{O}\left(M^{3}K\right)$
computational complexity, where the number of iterations $N_{I}$
for convergence are few (4-8), i.e. $N_{I}\ll\frac{KM^{2}}{d^{2}}$.
Similarly, non-iterative process has the similar complexity. Therefore,
the main computational intensive process is to compute the products
of matrices, e.g., $\mathbf{H}_{j}^{\dagger}\mathbf{H}_{j}$ or $\mathbf{V}_{j}^{\text{null}\dagger}\mathbf{H}_{j}^{\dagger}\mathbf{H}_{j}\mathbf{V}_{j}$. 

\subsection{Bounds on the harvesting Energy\label{subsec:Harvesting-Energy}}

It can be noted from the above analysis that not any trivial balanced
precoding can provide the better harvested energy. For the proposed
balanced precoding, the following bounds can be computed. 
\begin{lem}
\label{lem:bounds}Given the balanced precoders $\left\{ \mathbf{V}_{k}^{BAL},\forall k\right\} $
and IA precoders $\left\{ \mathbf{V}_{k},\forall k\right\} $ for
the channel $\left\{ \mathbf{H}_{kj},\forall k,j\right\} $, the total
harvested energy can be bounded as 
\begin{subequations}
\begin{gather}
\zeta\sum_{j=1}^{K}\frac{P_{j}}{d}\left[\left\Vert \mathbf{H}_{j}\mathbf{V}_{j}\right\Vert _{F}^{2}\left(1-\frac{z_{j}}{d}\right)+\left\Vert \mathbf{H}_{j}\mathbf{V}_{j}^{\text{n}}\right\Vert _{F}^{2}\left(\frac{z_{j}}{d}\right)\right]\\
\leq\sum_{k=1}^{K}Q_{k}(\bar{\rho}_{k},\mathbf{V}_{k}^{BAL})\leq\zeta\sum_{j=1}^{K}P_{j}\lambda_{j1},
\end{gather}
\end{subequations}
where the left and right equalities occur for $z_{j}=0$ and $z_{j}=z_{j}^{EH}$
respectively. 
\end{lem}
\begin{IEEEproof}
Proof is given in Appendix\ref{subsec:Proof-of-bounds}.
\end{IEEEproof}
The above result shows an improvement over \eqref{eq:max_EH-Qk} based
on the value of $z_{j}$, i.e., the balanced precoding guarantees
the better harvested energy than that achieved using just perfect
IA precoders. With the balanced precoding, the resultant rate loss
can be obtained from the upper bound in the Lemma \ref{lem:Rate-loss-basic}. 

With $\bar{\rho}_{k}=\bar{\rho},\forall k$ and with $\mathcal{CN}(0,1)$
entries in $\mathbf{H}_{kj}$, performing the expectation on both
the sides in the above equation gives 
\begin{subequations}
\begin{gather}
\zeta\bar{\rho}KN\sum_{j}P_{j}\approx\sum_{k=1}^{K}\mathbb{E}\left\{ Q_{k}(\bar{\rho})\right\} \\
\leq\zeta\bar{\rho}KNd\left(\frac{KN+d}{KNd+1}\right)^{2/3}\sum_{j=1}^{K}P_{j},
\end{gather}
\end{subequations}
where the left approximation is obtained assuming $\mathbb{E}\left\Vert \mathbf{H}_{j}\mathbf{V}_{j}^{BAL}\right\Vert _{F}^{2}$
$\approx$ $\bar{\rho}KNd$, and the right inequality is given by
$\mathbb{E}\left\{ \lambda_{j1}\right\} =\bar{\rho}KNd\left(\frac{KN+d}{KNd+1}\right)^{2/3}$
\cite{5770238}.

\section{Energy Harvesting with Feedback\label{sec:Energy-Harvesting-withFB}}

In the above formulation for EH, perfect IA precoder has been employed,
which is not the case in practice. In practice, to avail the precoder
at the transmitter side, either CSI or precoder is fed back in quantized
or analog form. In this section, considering precoder feedback, the
sum rate and energy harvesting terms are analyzed. Recall that the
trade off between these is characterized by the chordal distance.
Therefore, in the following analog precoder feedback scheme is provided,
followed by limited precoder feedback.

\subsection{Analog Feedback\label{subsec:Orthogonal-Transmission-of}}

In analog feedback, after the estimation of the reverse links, full
($M\times d$) precoder is sent back using analog transmission \cite{8379356}.
For orthogonal transmissions, destinations transmit simultaneously
in $Kd$ time slots respectively. After receiving the noisy precoder
information at the sources, the orthogonalization of the MMSE estimate
is performed to obtain the final precoder estimate.

From the results in \cite{8379356}, it can be observed that for
lower feedback SNR, the average chordal distance between the estimate
and IA precoder remains constant, which results in the sum rate loss
increasing with SNR (see Lemma \ref{lem:Rate-loss-basic}). On the
other hand, for medium to high SNR case, the chordal distance decreases
inversely proportional to SNR, which keeps the rate loss constant
for this SNR range. In IA scenarios, the medium to high SNR regime
is of more importance. In conjunction with energy harvesting, one
can note that the chordal distance is automatically set according
to the feedback SNR selected at the destinations. The conclusion of
this result is that analog feedback also helps in increasing energy
efficiency while maintaining linear sum rate scaling, that is, no-loss
of DoFs. To get the desired energy in harvesting, only the splitting
factor needs to be selected using the results in the previous section.

\subsection{Quantized Feedback\label{subsec:Precoder-Quantization-Perfect}}

Let the vector $\mathbf{b}=[b_{1},b_{2},\ldots,b_{K}]^{T}$ denote
the number of feedback bits allocated for each user. The corresponding
precoder quantization codebook of size $2^{b_{k}}$ is given as $\mathcal{C}(b_{k})=\{\mathbf{C}_{1}(b_{k}),\ldots,\mathbf{C}_{2^{b_{k}}}(b_{k})\}$
where each entry $\mathbf{C}_{i}(b_{k})$ is an $M\times d$ orthogonal
matrix such that $\mathbf{C}_{i}(b_{k})^{\dagger}\mathbf{C}_{i}(b_{k})=\mathbf{I}_{d}$.
The codebook $\mathcal{C}(b_{k})$ is considered known to all the
transmitters and receivers. The precoder matrix index (PMI) vector
is denoted as $\mathbf{q}=[q_{1},\ldots,q_{K}]^{T}$ with each $q_{k}$
representing an index from the codebook $\mathcal{C}(b_{k})$, i.e.
$1\leq q_{k}\leq2^{b_{k}}$, $\forall k$.

In the conventional method, when the perfect IA precoders are available,
each of the precoders is quantized using the chordal distance metric.
Let $\mathbf{q}_{CD}$ denote the PMI vector obtained using quantization
based on chordal distance. The $k^{th}$ index of $\mathbf{q}_{CD}$
is obtained as 
\begin{equation}
q_{CD,k}=\arg\min_{\mathbf{C}_{i}\in\mathcal{C}}d_{c}^{2}(\mathbf{V}_{k},\mathbf{C}_{i}),\:
\end{equation}
where $q_{CD,k}$ is the index of the closest codebook entry. This
technique incurs a low computational complexity. Improved precoder
feedback schemes can be seen in \cite{garg2015precoder,Garg2016quant}
which suggest that the sum rate can be improved for the same number
of quantization bits. The main point is to observe that the limited
feedback can increase the harvested energy since chordal distance
is non-zero. Note that the resultant chordal distance of quantized
precoders varies inversely proportional to the codebook size, i.e.,
as codebook size increases EH decreases and sum rate increases. Therefore,
given the codebook $\mathcal{C}(b_{k})$, the chordal distance can
be fixed as 
\begin{equation}
z=\mathbb{E}d_{c}^{2}(\mathbf{V}_{k},\mathbf{C}_{i})<2^{-\frac{b_{k}}{d(M-d)}},
\end{equation}
and splitting ratio can be varied to get the desired harvested energy.

\emph{Remark (SNR shift)}: For $z=0$, the splitting causes the noise
variance to change from $\sigma^{2}$ to $\sigma_{ID}^{2}$, which
causes the $\left(\frac{\sigma_{ID}^{2}}{\sigma^{2}}\right)_{dB}=\left(1+\frac{\delta^{2}}{\rho_{k}\sigma^{2}}\right)_{dB}$
shift in SNR without loosing linear sum rate scaling. It shows that
the splitting factor can be obtained for a given SNR shift or a given
constant rate loss. For example, to get only $3$ dB loss in sum rates,
$1+\frac{\delta^{2}}{\rho_{k}\sigma^{2}}=2$ or $\rho_{k}=1-\bar{\rho}_{k}=\frac{\delta^{2}}{\sigma^{2}}$.

\section{\label{sec:Simulation-Results}Simulation Results}

\subsection{Simulation settings}

The value of essential variables are given as follows: $\rho_{k}=\zeta=0.5,\forall k$,
$P_{k}=P$ and $z_{j}=z$. We consider two IA-feasible systems (a)
$(4\times4,2)^{3}$ and (b) $(5\times5,2)^{3}$. Each entry of $\mathbf{H}_{kj}$
is assumed to be distributed as $\mathcal{CN}(0,1)$. For balanced
precoding, the iterative process (ICD) is run for a maximum of $6$
iterations. We assume for all $j$, $z_{j}=z<\min_{j}z_{j}^{EH}$.
In the following figures, we compare different precoding strategies
given below. 
\begin{itemize}
\item (RAND) Random full rank precoder with orthogonal columns;
\item (MAX-EH) Harvested energy maximizing precoder;
\item (SSIA) Balanced Precoders from subspace IA method with $z=0,0.1,0.8$;
\item (MSEIA) Balanced precoder from MMSE based IA algorithm \cite{9097459}
with $z=0,0.1,0.8$;
\item (PQFB) Precoder obtained from quantized feedback from a given codebook
of size $8$ bits \cite{9097459}; 
\item Precoder acquired via analog precoder feedback (PAFB) \cite{8379356}
, and via analog CSI feedback (CSIAFB) \cite{Ayach2012} with similar
feedback transmission power as the forward one, $P_{f}=P$.
\end{itemize}

\subsection{Rate-energy region plots}

Given the precoders $\left\{ \mathbf{V}_{k},\forall k\right\} $,
the rate-energy region can be written as \cite{8241822,6623062},
$\mathcal{C}=$
\begin{equation}
\bigcup_{\boldsymbol{\rho}}\left\{ \left(R,Q\right):R\leq\sum_{k=1}^{K}R_{k}\left(\rho_{k},\mathbf{V}_{k}\right),Q\leq\sum_{k=1}^{K}Q_{k}\left(\rho_{k},\mathbf{V}_{k}\right)\right\} ,
\end{equation}
where $\boldsymbol{\rho}=\left[\rho_{1},\ldots,\rho_{K}\right]$ is
a vector of $K$ splitting factors. For a splitting noise variance
$\delta^{2}=0.1$, assuming $\rho_{k}=\rho,\forall k$ varying $\rho$
from $0$ to $1$, the parametric plots are drawn to illustrate rate-energy
regions \cite{8241822,6623062}. 

\begin{figure}
\centering\includegraphics[width=8.9cm]{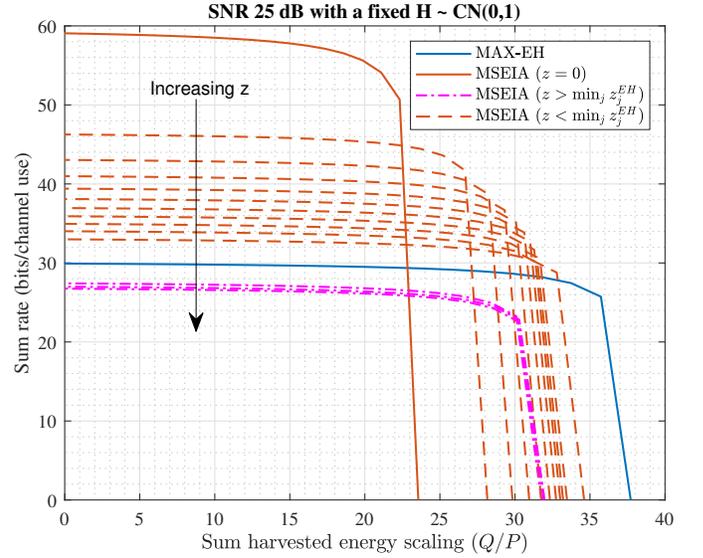}

\caption{Rate-energy plot for $\left(5\times5,2\right)^{3}$ system for iterative
CD based balanced precoding. \label{fig:Rate-energy-plot-for}}
\end{figure}
Figure \ref{fig:Rate-energy-plot-for} shows the sum rate versus the
total harvested energy plot for three types of precoders, viz., MSE-IA,
MAX-EH, and MSE-IA with balanced precoding with different values of
CD. It can be seen that MSE-IA region has higher sum rates and lower
energies, while the region for MAX-EH precoders has less sum rates
and higher energies. These plots represent two extreme ends of rate
and energy achievabililty. Next, for the balanced precoding with iterative
method, it can be observed that as $z$ increases, the rate decreases
and energy increases, when $z<\min_{j}z_{j}^{EH}$. When $z>\min_{j}z_{j}^{EH}$,
both rate and energy achieved are lower. Therefore, for the case of
$z>\min_{j}z_{j}^{EH}$, it is better to employ MAX-EH precoder than
IA-precoder. 

\begin{figure}
\centering\includegraphics[width=8.9cm]{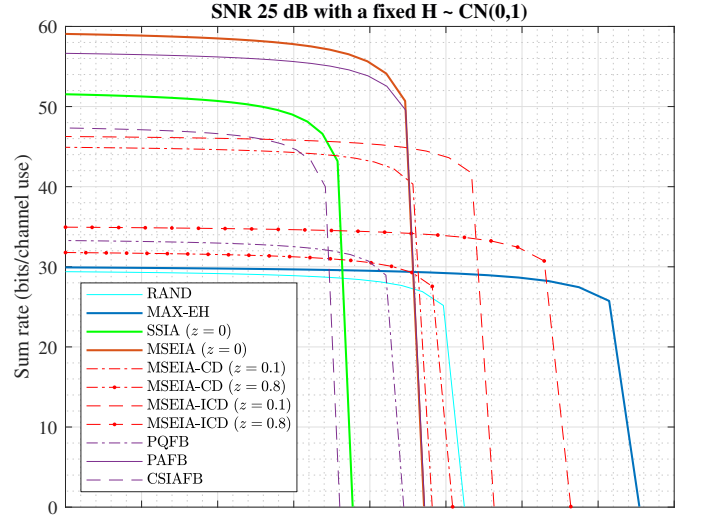}

\caption{Rate-energy plot for $\left(5\times5,2\right)^{3}$ system for different
precoding methods. \label{fig:Rate-energy-plot-for-ALL}}
\end{figure}
Figure \ref{fig:Rate-energy-plot-for-ALL} compares the same rate-energy
region for different precoding schemes. The following points can be
concluded from the figure. 
\begin{itemize}
\item $\mathcal{C}_{\text{RAND}}\subset\mathcal{C}_{\text{MAX-EH}}$ and
$\mathcal{C}_{\text{RAND}}\subset\mathcal{C}_{\text{MSEIA-ICD}}$:
Random precoders have worst rates. 
\item $\mathcal{C}_{\text{SSIA}}\subset\mathcal{C}_{\text{MSEIA}}$: Among
IA-methods, MSE based methods are better suited for both rate and
energy optimization. 
\item $\mathcal{C}_{\text{CSIAFB}}\subset\mathcal{C}_{\text{PAFB}}$ and
$\mathcal{C}_{\text{PQFB}}\subset\mathcal{C}_{\text{PAFB}}$: Analog
precoder feedback is better than both analog CSI feedback and precoder
quantized feedback. 
\item $\mathcal{C}_{\text{MSEIA-CD}}\subset\mathcal{C}_{\text{MSEIA-ICD}}$:
Iterative balanced precoding method provides better rate and energy
than that via non-iterative one. Therefore, in the following, iterative
method based precoding is considered for comparison. 
\end{itemize}

\subsection{Sum Rate and harvested energy versus SNRs}

\begin{figure}
\centering\includegraphics[width=8.9cm]{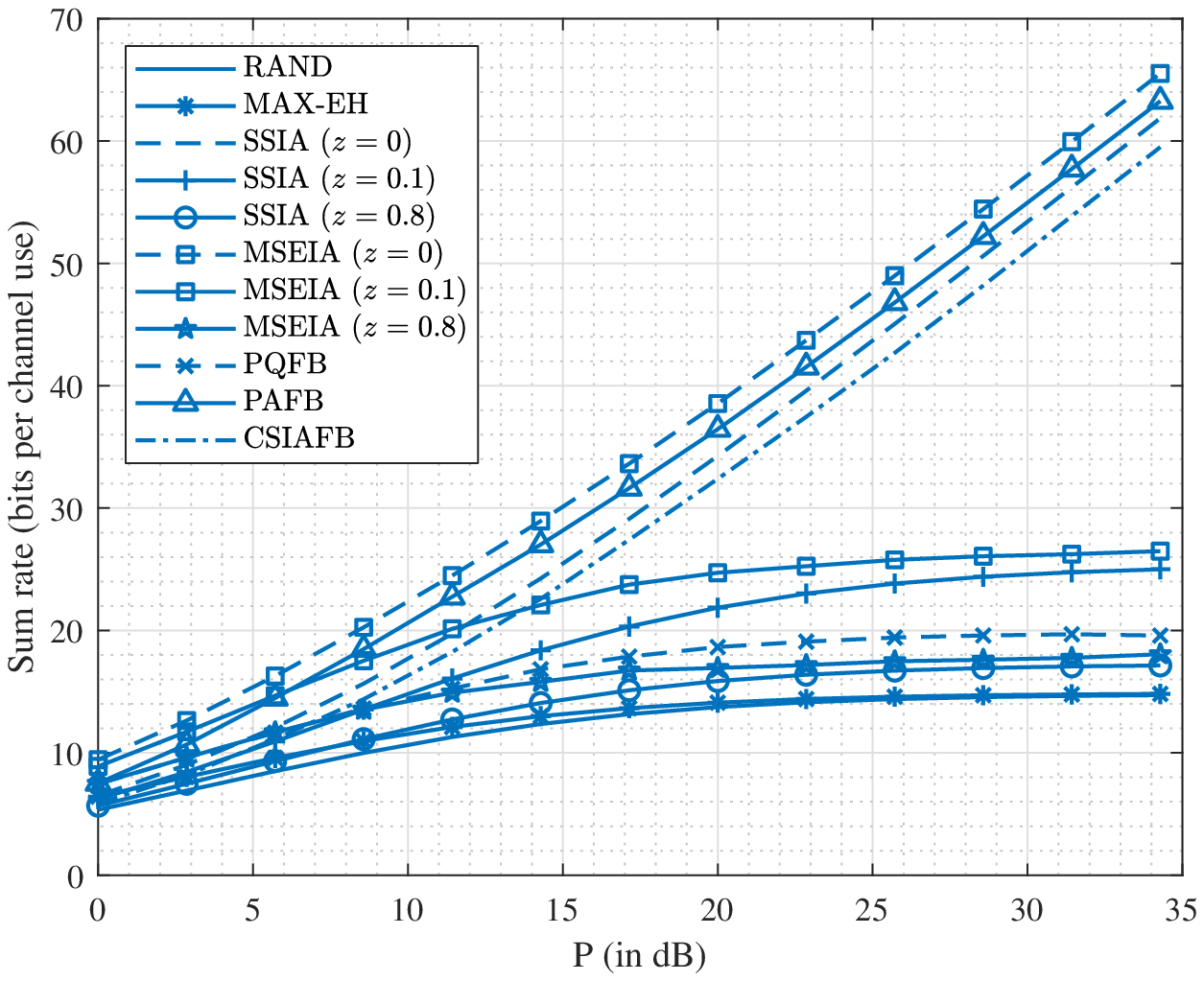}

\caption{Sum rates versus SNR with $(4\times4,2)^{3}$system. \label{fig:Figure-1-1}}
\end{figure}
\begin{figure}
\centering \includegraphics[width=8.9cm]{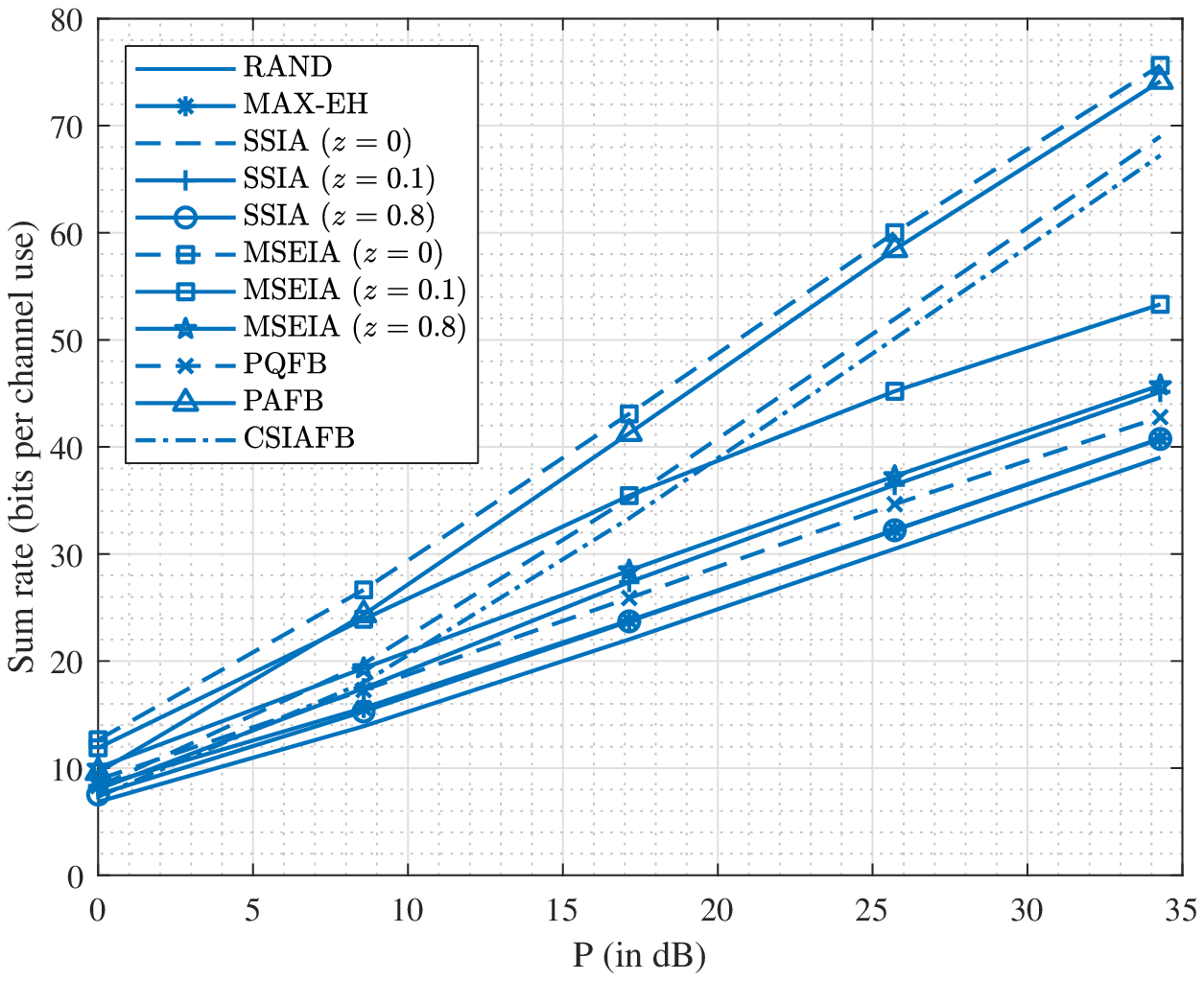}

\caption{Sum rates versus SNR with $(5\times5,2)^{3}$ system. \label{fig:Figure-1-1-1}}
\end{figure}

Figure \ref{fig:Figure-1-1}-\ref{fig:Figure-1-1-1} illustrate the
sum rates with respect to SNR for $(4\times4,2)^{3}$ and $(5\times5,2)^{3}$
systems, respectively. It can be observed that both SSIA and MSEIA
($z=0$) achieve linear sum rate scaling with SNR, while with $z>0$,
saturating sum rates are obtained at high SNR. As compared to $(4\times4,2)^{3}$
system, the saturation in sum rates starts at higher SNR for $(5\times5,2)^{3}$
system, since more spatial dimensions are available in $(5\times5,2)^{3}$
to grant diversity gains. For limited feedback with $8$ quantization
bits per precoder, similar rate losses can be seen due to saturation,
because to keep the rate loss constant number of bits need to be scaled
proportional to SNR \cite{9097459}. For analog feedback (AFB) schemes
(CSIAFB and PAFB), a constant rate loss can be observed at high SNR
regime, yielding the better performance than quantization schemes.
Figure \ref{fig:Figure-1-1-1} plots the similar trend for $(5\times5,2)^{3}$
system, except that higher sum rates are achieved due to more spatial
dimensions. The decreasing/saturating behavior of sum rates Figure
\ref{fig:Figure-1-1} can also be observed in Figure \ref{fig:Figure-1-1-1};
however, it requires much higher SNR in the $(5\times5,2)^{3}$ system
than the SNR in $(4\times4,2)^{3}$ system. 

\subsection{Rate-Energy performance versus chordal distance }

\begin{figure}
\centering\includegraphics[width=8.9cm]{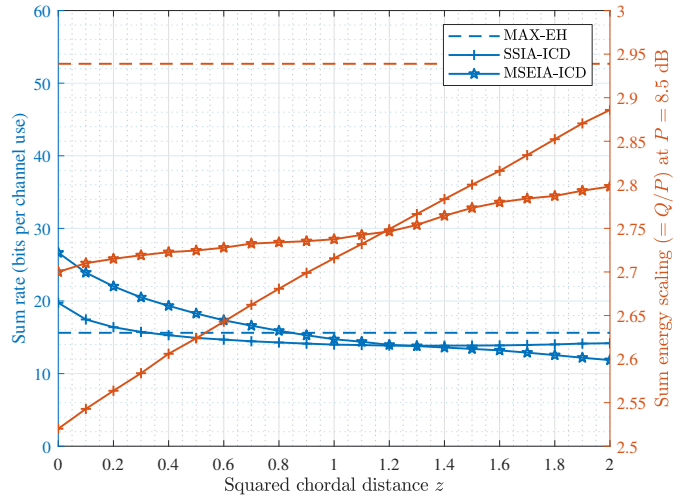}

\caption{Figure illustrates both the sum rate and sum harvested power variations
versus chordal distance for $(5\times5,2)^{3}$ at 25 dB SNR. \label{fig:Figure_vsZ}}
\end{figure}
Figure \ref{fig:Figure_vsZ} plots the sum rate (left-axis) and the
harvested energy (right-axis) versus the squared chordal distance
$z=d_{c}^{2}\left(\mathbf{V}_{j},\mathbf{V}_{j}^{BAL}\right),\forall j$
required for the balanced precoding with MSE-IA and SS-IA. It can
be seen that the sum rate decreases in a logarithmic manner as $z$
increases. This behavior has been analyzed in the Equation \eqref{eq:lemmarateloss}
for rate-loss upper bound. On the other hand, harvested energy is
increased, if $z$ is increased. Upto a certain value of $z$ (say
$z_{th}$), MSE-IA provides higher energies than that with SS-IA.
When $z>z_{th}$, SS-IA yields better energy output. Regarding the
sum rate intersection between MAX-EH and MSE-IA (or SS-IA), it is
the point when $z=\min_{j}z_{j}^{EH}$. When $z>\min_{j}z_{j}^{EH}$,
both rate and energy are lower than that of MAX-EH. Thus, it is better
to consider MAX-EH precoder beyond this intersection. Note that the
earlier intersection of SS-IA than MSE-IA is due to the fact that
MSE-IA provides better sum rates than SS-IA in general. 

\subsection{Energy scaling with SNRs}

\begin{figure}
\centering\includegraphics[width=8.9cm]{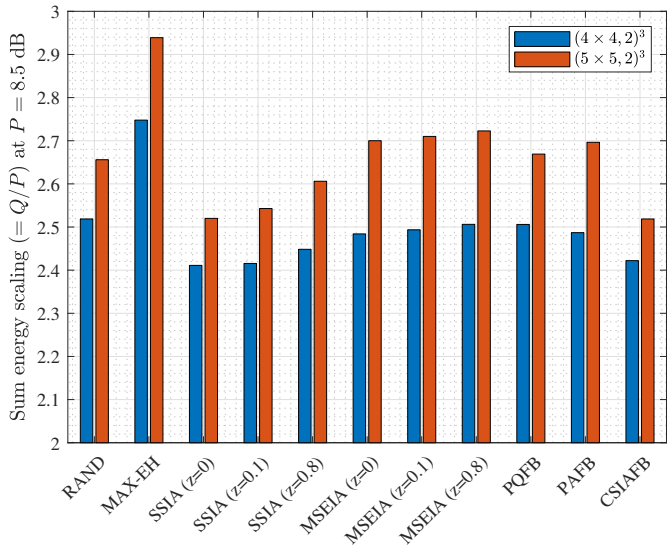}

\caption{Figure showing the sum harvested power scaling for different precoding
methods for $(4\times4,2)^{3}$ and $(5\times5,2)^{3}$ systems at
25 dB SNR. \label{fig:Figure-Ehscal}}
\end{figure}
The respective harvested energy scaling (with respect to transmit
power ($P$)) is illustrated in Figure \ref{fig:Figure-Ehscal} for
different precoding strategies. Max-EH precoding provides maximum
scaling. MSE-IA methods provide better scaling than SS-IA based ones.
Also, for rate-energy balanced precoding, increasing the chordal distance
shows increase in scaling. This result is also depicted in Figure
\ref{fig:Figure_vsZ}, which shows the sum rate and EH scaling variations
with respect to chordal distance. It can be seen that the energy scales
linearly with chordal distance. The most efficient method is PAFB,
where the chordal distance is selected automatically inversely proportional
to feedback SNR. It suggests to choose the chordal distance carefully
based on SNR and the required harvested energy constraint. 

\subsection{SER plots}

\begin{figure}
\centering \includegraphics[width=8.9cm]{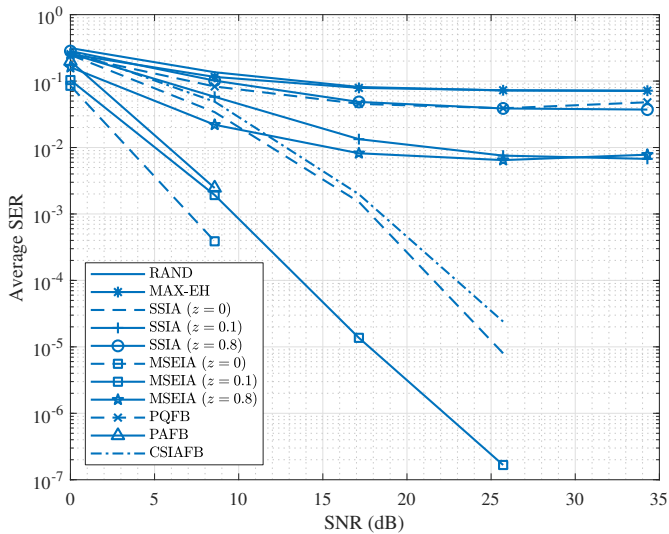}

\caption{Figure depicts the average QPSK symbol error rate (SER) with respect
to SNR for $(5\times5,2)^{3}$. \label{fig:Figure-ser}}
\end{figure}
Figure \ref{fig:Figure-ser} depicts the average SER plots with QPSK
modulation for $(5\times5,2)^{3}$. It can be seen that perfect IA
precoders ($z=0$) achieve the minimum SER, while with $z=0.1,0.8$,
the SER saturates. For quantization based methods, the SER can be
seen to be higher. Among the feedback schemes, PAFB methods can be
seen to provide a significantly better SER, close to perfect MSE-IA
scheme. More importantly, PAFB methods also yield linear sum rate
scaling and EH scaling approximately to MSE-IA ($z=0.1$), which shows
the effectiveness of PAFB schemes.

\section{Conclusion\label{sec:Conclusion}}

In this paper, we have provided a low-complexity systematic balanced
precoding method towards getting the improved trade-off between sum
rate and harvested energy. First, the precoder that achieves maximum
harvested energy has been computed. Thereafter, utilizing the CD decomposition,
we have systematically derived and analyzed the proposed precoder
for SWIPT trade-off. The lower and upper bounds on harvested energy
for this construction have been obtained. Due to the dependence on
CD, the relations to the analog and quantized feedback schemes have
been discussed. Simulation results show that MSE based methods are
better for SWIPT than subspace alignment method or leakage minimization
algorithm. Among feedback schemes, PAFB has shown to improve sum rates
without loosing the linear scaling (with SNR) as well as improved
harvested energy. 

The future work is to investigate the proposed schemes in case of
hybrid precoding with non-linear hardware impairments and constraints.

\appendices{}

\section*{Appendices}

\subsection{Proof of CD decomposition\label{sec:Proof-of-CD-1}}

Consider two $M\times d$ orthonormal matrices $\mathbf{V},\hat{\mathbf{V}}$
such that $\mathbf{V}^{\dagger}\mathbf{V}=\hat{\mathbf{V}}^{\dagger}\hat{\mathbf{V}}=\mathbf{I}_{d}$.
Its left null space of size $M\times M-d$ can be represented as $\hat{\mathbf{V}}_{j}^{\text{null}}=\text{null}(\hat{\mathbf{V}}_{j})$.
Then, we can write 
\begin{align}
\mathbf{V} & =\hat{\mathbf{V}}\hat{\mathbf{V}}^{\dagger}\mathbf{V}+\left(\mathbf{I}_{M}-\hat{\mathbf{V}}\hat{\mathbf{V}}^{\dagger}\right)\mathbf{V}\\
 & =\hat{\mathbf{V}}\underbrace{\hat{\mathbf{V}}^{\dagger}\mathbf{V}}_{=\mathbf{X}\mathbf{Y}}+\hat{\mathbf{V}}^{\text{null}}\underbrace{\hat{\mathbf{V}}^{\text{null}\dagger}\mathbf{V}}_{=\mathbf{S}\mathbf{Z}}
\end{align}
where the last equation is obtained by the QR-decomposition such that
$\mathbf{X}$ and $\mathbf{S}$ are $d\times d$ and $M-d\times d$
orthonormal matrices respectively. It verifies $d_{c}^{2}(\mathbf{V},\hat{\mathbf{V}})=d-\|\hat{\mathbf{V}}^{\dagger}\mathbf{V}\|_{F}^{2}=d-\text{tr}(\mathbf{Y}^{\dagger}\mathbf{Y})=\text{tr}(\mathbf{Z}^{\dagger}\mathbf{Z})$.
Note that $\mathbf{X}\mathbf{Y}\in\mathbb{C}^{d\times d}$ is independent
of $\hat{\mathbf{V}}\in\mathbb{C}^{M\times d}$, since $\mathbf{XY}$
is a projection to a lower dimension space. Also, the factors $\mathbf{X}$
and $\mathbf{Y}$ are independent, since $\mathbf{X}$ represents
the basis of $\hat{\mathbf{V}}^{\dagger}\mathbf{V}$ and the basis
are not unique. Using similar facts, the matrices $\mathbf{S}$ and
$\mathbf{Z}$ are also independent. For more details, visit \cite{Ravindran2008}. 

\subsection{\label{subsec:Proof-of-RLUB}Proof of Lemma \ref{lem:Rate-loss-basic}}
\begin{IEEEproof}
In literature, it is known that the rate loss is upper bounded by
the interference terms \cite{Ravindran2008,Ayach2012}. Thus, the
rate loss bound can be expressed as 
\begin{align}
\Delta R_{k} & \leq\log_{2}\left|\mathbf{I}_{d}+\frac{P}{d\sigma_{ID}^{2}}\sum_{j\neq k}\mathbb{E}\text{\ensuremath{\underbar{\ensuremath{\mathbf{H}}}}}_{kj}^{\dagger}\hat{\mathbf{V}}_{j}\hat{\mathbf{V}}_{j}^{\dagger}\text{\ensuremath{\underbar{\ensuremath{\mathbf{H}}}}}_{kj}\right|,
\end{align}
where $\text{\ensuremath{\underbar{\ensuremath{\mathbf{H}}}}}_{kj}^{\dagger}=\mathbf{U}_{k}^{\dagger}\mathbf{H}_{kj}$.
Using Lemma \ref{lem:MAT-CD-decomp-1}, we can write  
\begin{align}
\text{\ensuremath{\underbar{\ensuremath{\mathbf{H}}}}}_{kj}^{\dagger}\hat{\mathbf{V}}_{j} & =\text{\ensuremath{\underbar{\ensuremath{\mathbf{H}}}}}_{kj}^{\dagger}\mathbf{V}_{j}\mathbf{X}_{j}\mathbf{Y}_{j}+\text{\ensuremath{\underbar{\ensuremath{\mathbf{H}}}}}_{kj}^{\dagger}\mathbf{S}_{j}\mathbf{Z}_{j}\\
 & =\text{\ensuremath{\underbar{\ensuremath{\mathbf{H}}}}}_{kj}^{\dagger}\mathbf{S}_{j}\mathbf{Z}_{j}
\end{align}
where $\text{\ensuremath{\underbar{\ensuremath{\mathbf{H}}}}}_{kj}^{\dagger}\mathbf{V}_{j}=0$
for interference alignment.  To make $\text{\ensuremath{\underbar{\ensuremath{\mathbf{H}}}}}_{kj}$
orthonormal, let $\tilde{\mathbf{H}}_{kj}=\text{\ensuremath{\underbar{\ensuremath{\mathbf{H}}}}}_{kj}\mathbf{W}_{kj}\mathbf{\Lambda}_{kj}^{-1/2}$
such that, $\tilde{\mathbf{H}}_{kj}^{\dagger}\tilde{\mathbf{H}}_{kj}=\mathbf{I}_{d}$,
where $\text{\ensuremath{\underbar{\ensuremath{\mathbf{H}}}}}_{kj}^{\dagger}\text{\ensuremath{\underbar{\ensuremath{\mathbf{H}}}}}_{kj}=\mathbf{W}_{kj}\mathbf{\Lambda}_{kj}\mathbf{W}_{kj}^{\dagger}$
be the eigenvalue decomposition. The above decomposition is similar
to SVD, $\text{\ensuremath{\underbar{\ensuremath{\mathbf{H}}}}}_{kj}=\tilde{\mathbf{H}}_{kj}\mathbf{\Lambda}_{kj}^{1/2}\mathbf{W}_{kj}^{\dagger}$,
where $\tilde{\mathbf{H}}_{kj}$, $\mathbf{W}_{kj}$, $\mathbf{\Lambda}_{kj}$
are independent of each other. Since $\mathbf{S}_{j}$ and $\mathbf{Z}_{j}$
are independent as well, the following product can be simplified as
\begin{align}
 & \mathbb{E}\text{\ensuremath{\underbar{\ensuremath{\mathbf{H}}}}}_{kj}^{\dagger}\mathbf{S}_{j}\mathbf{Z}_{j}\mathbf{Z}_{j}^{\dagger}\mathbf{S}_{j}^{\dagger}\text{\ensuremath{\underbar{\ensuremath{\mathbf{H}}}}}_{kj}\\
\stackrel{(a)}{=} & \mathbb{E}\mathbf{W}_{kj}(\mathbf{\Lambda}_{kj}^{1/2})^{\dagger}\tilde{\mathbf{H}}_{kj}^{\dagger}\mathbf{S}_{j}\mathbf{Z}_{j}\mathbf{Z}_{j}^{\dagger}\mathbf{S}_{j}^{\dagger}\tilde{\mathbf{H}}_{kj}\mathbf{\Lambda}_{kj}^{1/2}\mathbf{W}_{kj}^{\dagger}\\
\stackrel{(b)}{=} & \mathbb{E}\mathbf{W}_{kj}(\mathbf{\Lambda}_{kj}^{1/2})^{\dagger}\tilde{\mathbf{H}}_{kj}^{\dagger}\mathbf{S}_{j}\left[\mathbb{E}\mathbf{Z}_{j}\mathbf{Z}_{j}^{\dagger}\right]\mathbf{S}_{j}^{\dagger}\tilde{\mathbf{H}}_{kj}\mathbf{\Lambda}_{kj}^{1/2}\mathbf{W}_{kj}^{\dagger}\\
\stackrel{(c)}{=} & \frac{z}{d}\mathbb{E}\mathbf{W}_{kj}(\mathbf{\Lambda}_{kj}^{1/2})^{\dagger}\left[\mathbb{E}\tilde{\mathbf{H}}_{kj}^{\dagger}\mathbf{S}_{j}\mathbf{S}_{j}^{\dagger}\tilde{\mathbf{H}}_{kj}\right]\mathbf{\Lambda}_{kj}^{1/2}\mathbf{W}_{kj}^{\dagger}\\
= & \frac{z}{d}\frac{d}{M-d}\mathbb{E}\mathbf{W}_{kj}\mathbf{\Lambda}_{kj}\mathbf{W}_{kj}^{\dagger}\\
= & \frac{z}{M-d}\mathbb{E}\text{\ensuremath{\underbar{\ensuremath{\mathbf{H}}}}}_{kj}^{\dagger}\text{\ensuremath{\underbar{\ensuremath{\mathbf{H}}}}}_{kj}=\frac{z}{M-d}\bar{\Lambda}_{kj}=\frac{zM\mathbf{I}_{d}}{M-d}
\end{align}
where in (a), the decomposition of $\text{\ensuremath{\underbar{\ensuremath{\mathbf{H}}}}}_{kj}=\tilde{\mathbf{H}}_{kj}\mathbf{\Lambda}_{kj}^{1/2}\mathbf{W}_{kj}^{\dagger}$
has been substituted; in (b), the expectation on $\mathbf{Z}_{j}$
is carried out, which is approximated to be $\frac{1}{d}\mathbb{E}tr(\mathbf{Z}_{j}\mathbf{Z}_{j}^{\dagger})=\frac{z_{j}}{d}$
as \cite[App. B]{Ravindran2008}, where $z_{j}$ is the expected chordal
distance; in (c), the quantity $\tilde{\mathbf{H}}_{kj}^{\dagger}\mathbf{S}_{j}$
is matrix Beta distributed $BETA(d,M-2d)$, which has mean of $\frac{d}{M-d}$;
and we write $\bar{\Lambda}_{kj}=\mathbb{E}\text{\ensuremath{\underbar{\ensuremath{\mathbf{H}}}}}_{kj}^{\dagger}\text{\ensuremath{\underbar{\ensuremath{\mathbf{H}}}}}_{kj}=\mathbb{E}_{H}\mathbf{U}_{k}^{\dagger}\mathbf{H}_{kj}\mathbf{H}_{kj}^{\dagger}\mathbf{U}_{k}\leq\mathbf{U}_{k}^{\dagger}\mathbb{E}_{H}\mathbf{H}_{kj}\mathbf{H}_{kj}^{\dagger}\mathbf{U}_{k}=M\mathbf{U}_{k}^{\dagger}\mathbf{U}_{k}=M\mathbf{I}_{d}$,
where the inequality occurs since $\mathbf{U}_{k}$ and $\mathbf{H}_{kj}$
are dependent. Thus, the rate loss bound expression can be given as
\begin{align}
\Delta R_{k}< & \log_{2}\left|\mathbf{I}_{d}+\frac{P}{d\sigma_{ID}^{2}}\sum_{j\neq k}\frac{z_{j}M\mathbf{I}_{d}}{M-d}\right|\\
= & d\log_{2}\left(1+\frac{P}{\sigma_{ID}^{2}}\frac{M}{d(M-d)}\sum_{j\neq k}z_{j}\right).
\end{align}
\end{IEEEproof}

\subsection{Proof of Lemma \ref{lem:bounds}\label{subsec:Proof-of-bounds}}
\begin{IEEEproof}
The inequality in the upper bound comes from \eqref{eq:max_EH-Vj}
as $\frac{1}{d}\sum_{i=1}^{d}\left\Vert \mathbf{H}_{j}\mathbf{V}_{j}^{BAL}\right\Vert _{F}^{2}\leq\frac{1}{d}\sum_{i=1}^{d}\lambda_{ji}\leq\lambda_{j1},\forall j$,
where the equality occurs when $z_{j}=z_{j}^{EH},\forall j$, and
the second inequality is due to the fact that the average of $d$-values
is less than the the maximum of them. 

The inequality of the lower bound can be derived from the CD decomposition,
where the equality occurs, when $z_{j}=0,\forall j$. For the proposed
balanced precoder with the optimum values of $\mathbf{X}_{j}^{*},\mathbf{Y}_{j}^{*},\mathbf{S}_{j}^{*}$
and $\mathbf{Z}_{j}^{*}$, we can write 
\begin{align*}
 & \left\Vert \mathbf{H}_{j}\mathbf{V}_{j}^{BAL}\right\Vert _{F}^{2}\\
 & =\left\Vert \mathbf{H}_{j}\mathbf{V}_{j}\mathbf{X}_{j}^{*}\mathbf{Y}_{j}^{*}+\mathbf{H}_{j}\mathbf{V}_{j}^{\text{null}}\mathbf{S}_{j}^{*}\mathbf{Z}_{j}^{*}\right\Vert _{F}^{2}\\
 & \stackrel{(a)}{\geq}\left\Vert \mathbf{H}_{j}\mathbf{V}_{j}\mathbf{X}_{j}^{*}\sqrt{1-\frac{z_{j}}{d}}+\mathbf{H}_{j}\mathbf{V}_{j}^{\text{null}}\mathbf{S}_{j}^{*}\sqrt{\frac{z_{j}}{d}}\right\Vert _{F}^{2}\\
 & \stackrel{(b)}{\geq}\left\Vert \mathbf{H}_{j}\mathbf{V}_{j}\mathbf{X}_{j}^{*}\right\Vert _{F}^{2}\left(1-\frac{z_{j}}{d}\right)+\left\Vert \mathbf{H}_{j}\mathbf{V}_{j}^{\text{null}}\mathbf{S}_{j}^{*}\right\Vert _{F}^{2}\left(\frac{z_{j}}{d}\right)\\
 & \stackrel{(c)}{\geq}\left\Vert \mathbf{H}_{j}\mathbf{V}_{j}\right\Vert _{F}^{2}\left(1-\frac{z_{j}}{d}\right)+\left\Vert \mathbf{H}_{j}\mathbf{V}_{j}^{\text{n}}\right\Vert _{F}^{2}\left(\frac{z_{j}}{d}\right),
\end{align*}
where in $(a)$, the maximum value of norm is upper bounded by trivial
selection $\mathbf{Z}_{j}=\mathbf{I}\sqrt{1-\frac{z_{j}}{d}}$; in
$(b)$, we employ the fact that the trace value in the above norm-expansion
is non-negative for the proposed scheme, as mentioned in the proposition
\ref{prop:trace-val-noneg}; and in $(c)$, the specific $d$-dimensional
null space $\left(\mathbf{V}_{j}^{\text{null}}\mathbf{S}_{j}^{*}\right)$
can be replaced with any other $d$-dimensional null space $\mathbf{V}_{j}^{\text{n}}=\mathbf{V}_{j}^{\text{null}}\mathbf{S}_{j}^{*}\in\mathcal{G}_{M,d}$
of $\mathbf{V}_{j}$. 
\end{IEEEproof}
\bibliographystyle{IEEEtran}
\addcontentsline{toc}{section}{\refname}\bibliography{avinash1}

\begin{IEEEbiography}[{\includegraphics[width=1in]{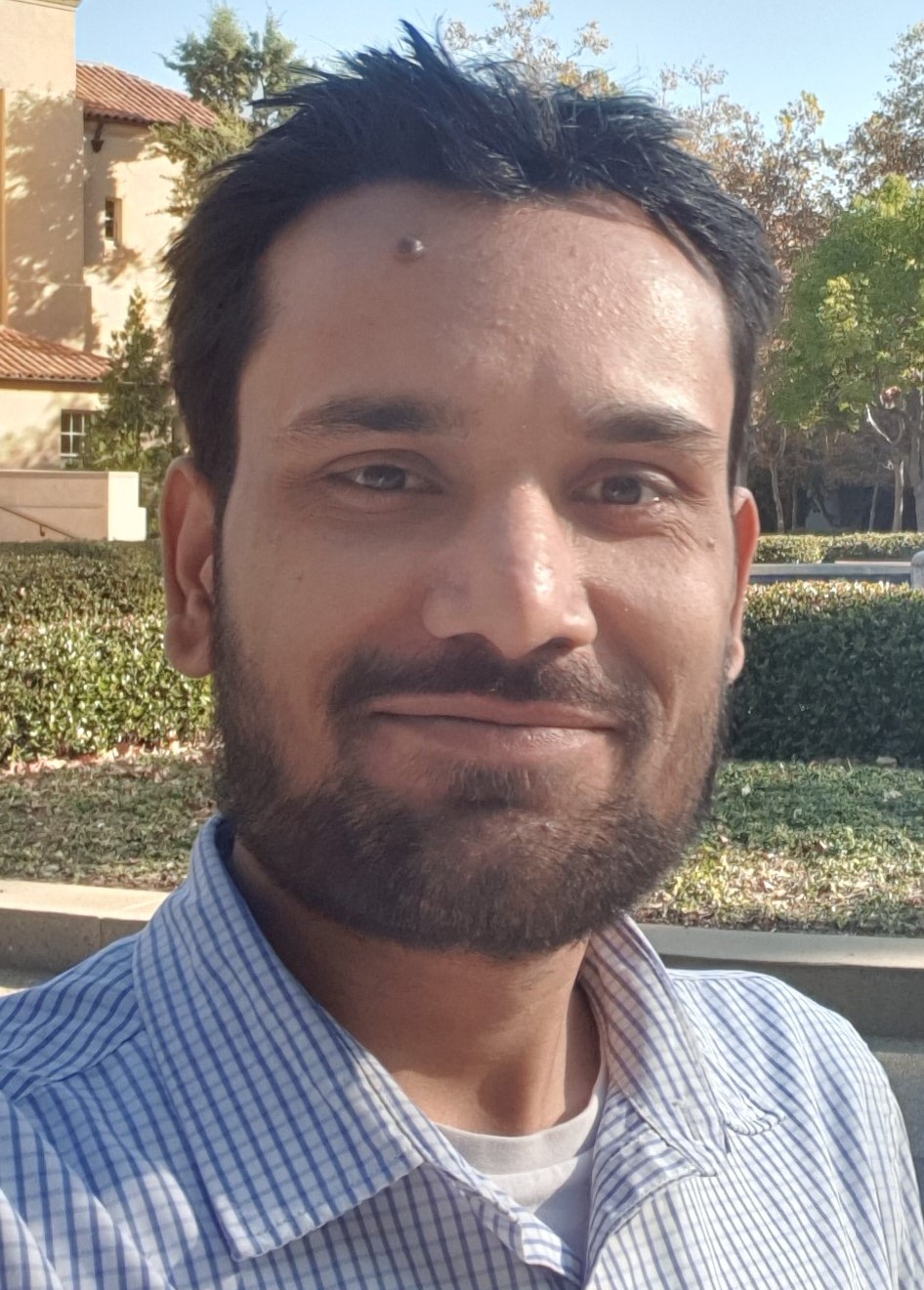}}]{Navneet Garg}
 received the B.Tech. degree in electronics and communication engineering
from College of Science \& Engineering, Jhansi, India, in 2010, and
the M.Tech. degree in digital communications from ABV-Indian Institute
of Information Technology and Management, Gwalior, in 2012. He has
completed the Ph.D. degree in June 2018 from the department of electrical
engineering at the Indian Institute of Technology Kanpur, India. From
July 2018-Jan 2019, he visited The University of Edinburgh, UK. From
February 2019-2020, he is employed as a research associate in Heriot-Watt
university, Edinburgh, UK. Since February 2020, he is working as a
research associate in The University of Edinburgh, UK. His main research
interests include wireless communications, signal processing, optimization,
and machine learning.

\end{IEEEbiography}

\begin{IEEEbiography}[{\includegraphics[width=1in]{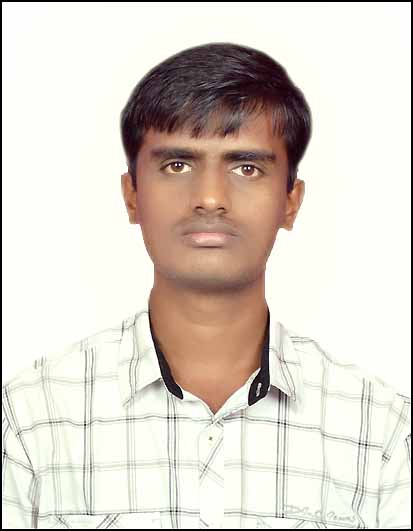}}]{Avinash Rudraksh}
 received B.Tech degree in Electronics \& Telecommunication from
National Institute of Technology Raipur, India, in 2012 and M.Tech
in Signal Processing and Wireless Networks from Institute of Information
Technology Kanpur, India, in 2018. He is currently working as a software
developer. His research interests include  wireless communications
and signal processing.

\end{IEEEbiography}

\begin{comment}
\end{comment}

\begin{IEEEbiography}[{\includegraphics[width=1in]{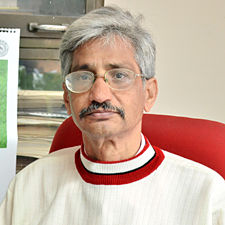}}]{Govind Sharma}
 received the B.Tech. and M.Tech. degrees in Department of Electrical
Engineering from Indian Institute of Technology Kanpur, India, in
1979 and 1981, and the Ph.D. degree from University of Southern California
(USC), in 1984. Since then, he has been a Professor with Department
of Electrical Engineering at Indian Institute of Technology Kanpur,
India. His general interests span the areas of signal processing and
communications, detection and estimation theory, etc. 

\end{IEEEbiography}

\begin{comment}
\end{comment}

\begin{IEEEbiography}[{\includegraphics[angle=270,width=1in]{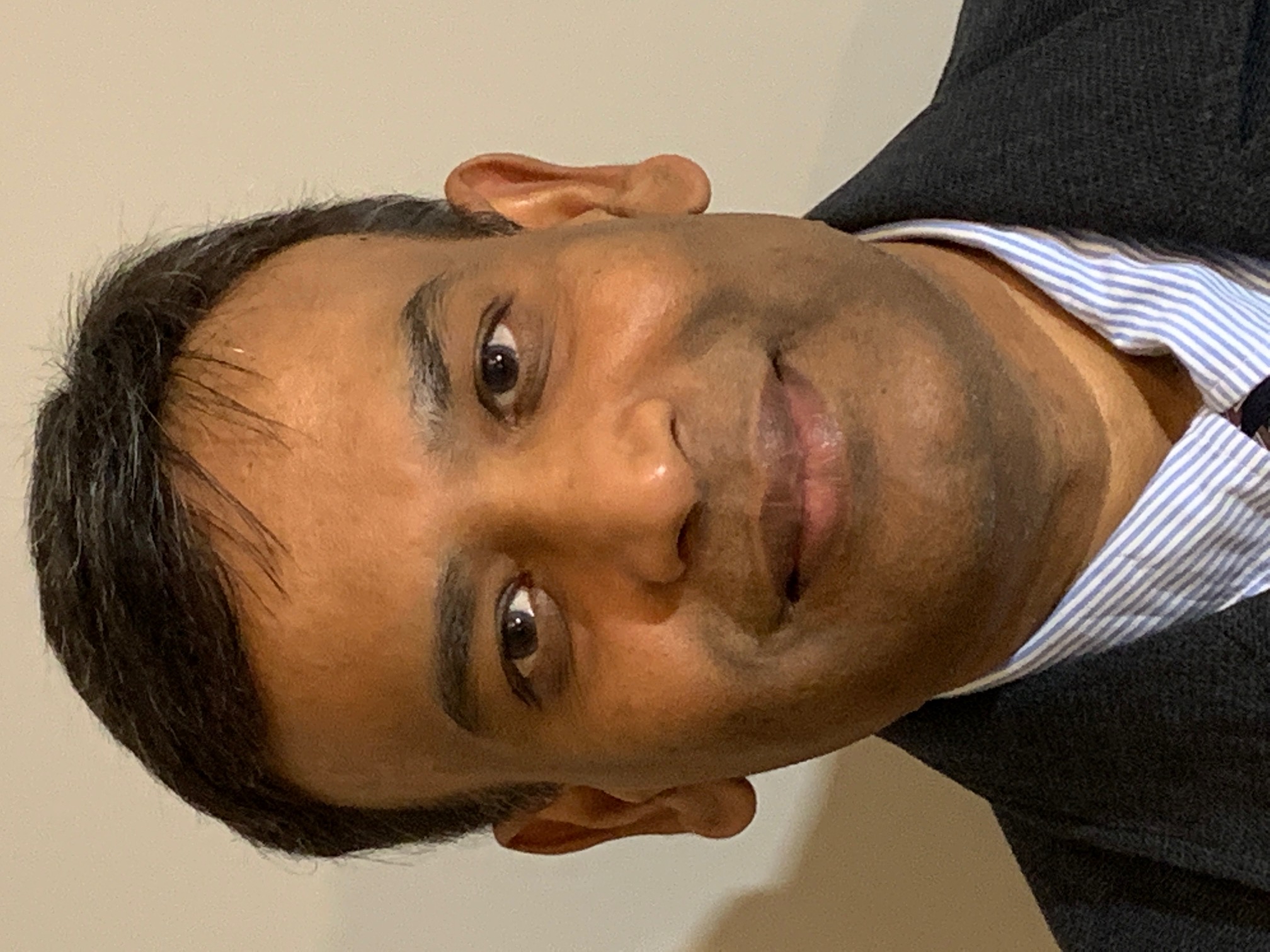}}]{Tharmalingam Ratnarajah}
 is currently with the Institute for Digital Communications, the
University of Edinburgh, Edinburgh, UK, as a Professor in Digital
Communications and Signal Processing. He was a Head of the Institute
for Digital Communications during 2016-2018. His research interests
include signal processing and information theoretic aspects of beyond
5G wireless networks, full-duplex radio, mmWave communications, random
matrices theory, interference alignment, statistical and array signal
processing and quantum information theory. He has published over 400
publications in these areas and holds four U.S. patents. He has supervised
16 PhD students and 21 post-doctoral research fellows and raised \$11+
million USD of research funding. He was the coordinator of the EU
projects ADEL (3.7M €) in the area of licensed shared access for 5G
wireless networks, HARP (4.6M €) in the area of highly distributed
MIMO, as well as EU Future and Emerging Technologies projects HIATUS
(3.6M €) in the area of interference alignment and CROWN (3.4M €)
in the area of cognitive radio networks. Dr Ratnarajah was an associate
editor IEEE Transactions on Signal Processing, 2015-2017 and Technical
co-chair, The 17th IEEE International workshop on Signal Processing
advances in Wireless Communications, Edinburgh, UK, 3-6, July 2016.
Dr Ratnarajah is a Fellow of Higher Education Academy (FHEA).
\end{IEEEbiography}

\begin{comment}
\end{comment}

\end{document}